\documentclass[journal]{IEEEtran}
\usepackage{amsmath,amsfonts}
\usepackage{algorithmicx,algorithm}
\usepackage{algpseudocode}  
\usepackage{array}
\usepackage[caption=false,font=normalsize,labelfont=sf,textfont=sf]{subfig}
\usepackage{textcomp}
\usepackage{stfloats}
\usepackage{url}
\usepackage{verbatim}
\usepackage{graphicx}
\usepackage{cite}
\usepackage{epstopdf}
\newtheorem{theorem}{Theorem}
\newtheorem{lemma}{Lemma}
\DeclareMathOperator{\tr}{tr}

\begin{document}

\title{Performance Analysis of Local Partial MMSE Precoding Based User-Centric Cell-Free Massive MIMO Systems and Deployment Optimization}

\author{Peng Jiang, Jiafei Fu,~\IEEEmembership{Student Member,~IEEE,} Pengcheng Zhu,~\IEEEmembership{Member,~IEEE,} Yan Wang, Jiangzhou Wang,~\IEEEmembership{Fellow,~IEEE,} Xiaohu You,~\IEEEmembership{Fellow,~IEEE}

\thanks{This work has been submitted to the IEEE for possible publication.  Copyright may be transferred without notice, after which this version may no longer be accessible.}
\thanks{This work was supported by the National Key R\&D Program of China under Grant 2021YFB2900300, the National Natural Science Foundation of China under Grant 62171126. \emph{(Corresponding author: Pengcheng Zhu.)}}
\thanks{Peng Jiang, Jiafei Fu and Yan Wang are with the National Mobile Communications Research Laboratory, Southeast University, Nanjing 210096, China (e-mail: jiangpeng98@seu.edu.cn; fujfei@seu.edu.cn; p.zhu@seu.edu.cn; yanwang@seu.edu.cn).}
\thanks{Pengcheng Zhu, Jiangzhou Wang and Xiaohu You are with the National Mobile Communications Research Laboratory, Southeast University, Nanjing 210096, China, and also with Purple Mountain Laboratories, Nanjing 211111, China (e-mail: p.zhu@seu.edu.cn; j.z.wang@kent.ac.uk; xhyu@seu.edu.cn).}
}

\markboth{Journal of \LaTeX\ Class Files,~Vol.~14, No.~8, August~2021}%
{Shell \MakeLowercase{\textit{et al.}}: A Sample Article Using IEEEtran.cls for IEEE Journals}


\maketitle

\begin{abstract}
Cell-free massive multiple-input multiple-output (MIMO) systems, leveraging tight cooperation among wireless access points, exhibit remarkable signal enhancement and interference suppression capabilities, demonstrating significant performance advantages over traditional cellular networks. 
This paper investigates the performance and deployment optimization of a user-centric scalable cell-free massive MIMO system with imperfect channel information over correlated Rayleigh fading channels.
Based on the  large-dimensional random matrix theory, this paper presents the deterministic equivalent of the ergodic sum rate for this system when applying the local partial minimum mean square error (LP-MMSE) precoding method, along with its derivative with respect to the channel correlation matrix.
Furthermore, utilizing the derivative of the ergodic sum rate, this paper designs a Barzilai-Borwein based gradient descent method to improve system deployment.
Simulation experiments demonstrate that under various parameter settings and large-scale antenna configurations, the deterministic equivalent of the ergodic sum rate accurately approximates the Monte Carlo ergodic sum rate of the system. 
Furthermore, the deployment optimization algorithm effectively enhances the ergodic sum rate of this system by optimizing the positions of access points.
\end{abstract}


\section{Introduction}
\IEEEPARstart{I}{n recent} years, massive multiple-input multiple-output (MIMO) technology has played a crucial role in the fifth generation (5G) communication networks, significantly driving leaps in the performance and efficiency of communication systems \cite{SHAFI20175JSAC,LARSSON2014MCOM,BJORNSON2016MCOM,Hoydis2013JSAC}. Within this technological framework, the deployment of a large number of antennas not only significantly enhances signal strength through beamforming techniques but also introduces additional spatial degrees of freedom that notably improve interference suppression within cells when serving multiple users \cite{Lu2014JSTSP,MARZETTA2010TWC}.
However, despite the optimization of system performance achieved by traditional cellular-based massive MIMO systems through frequency reuse and inter-cell interference management strategies, their inefficient frequency reuse mechanisms and inter-cell interference issues still pose constraints on the overall performance of communication systems.
To overcome these challenges and further drive the evolution of communication networks, cell-free massive MIMO systems are regarded as a highly promising networking solution \cite{NGO2017TWC,AKYILDIZ2020ACCESS,You2020SCIS}.

In an ideal cell-free system, all access points (APs) are centrally managed by a computation center, which uniformly handles operations such as resource allocation and precoding design, eliminating the concept of traditional cells \cite{NAYEBI2017TWC,ZHANG2019ACCESS,NGO2017TWC,HU2019TCOMM}. 
This system leverages the collaboration of numerous geographically distributed APs to reduce the average path loss for users, enhance the potential gain in MIMO signal strength, and provide the system with increased spatial degrees of freedom and spatial multiplexing gains, enabling stronger signals and reduced multi-user interference \cite{NGO2017TWC}.
However, implementing such an ideal system necessitates a powerful computation center capable of storing and processing all computations within the system. 
As the communication system expands and incorporates more APs, antennas, and users, the required storage capacity, computational complexity, and fronthaul capacity will increase linearly or even faster. 
This scalability challenge poses practical constraints for real-world deployment \cite{BUZZI2019TWC,BJORNSON2020TWC,AMMAR2022COMST,LI2023JIOT}. 
Therefore, achieving scalability becomes crucial for the practical implementation of cell-free massive MIMO systems, requiring that the resources such as computational load and backhaul capacity do not exceed the capacity of any node in the system as it expands indefinitely \cite{BJORNSON2020TWC}.

In the context of cell-free MIMO networks, adopting a user-centric service strategy has emerged as a feasible approach for deployment. 
The core of this system involves defining a limited range of APs around each user to form a cooperative cluster for service. 
This approach ensures that each user is associated only with nearby APs, and each AP focuses on serving a limited group of users in its vicinity, collaborating with a small number of adjacent APs to form an efficient service network \cite{BUZZI2019TWC,AMMAR2022COMST,LI2023JIOT,ZHENG2023TVT}.
Under this design framework, the computational load and backhaul resource consumption of APs are strictly limited to the actual demands of users within their direct service areas. 
Consequently, as the coverage area of the entire communication network expands, the resource requirements of each AP do not escalate indefinitely. 
Additionally, due to path loss, signals from distant users weaken significantly, posing minimal impact on the performance of nearby APs. 
This implies that even with cooperation limited to smaller geographical areas, the system can still approach the efficient performance level achievable under ideal unlimited-range cooperation \cite{AMMAR2022COMST}.

In the context of scalable system design, traditional high-performance precoding techniques, such as zero-forcing (ZF) and minimum mean square error (MMSE), which rely on global channel information, exhibit limitations. 
To address these limitations, precoding schemes designed for scalability have emerged. 
These schemes do not require frequent exchange of channel state information (CSI), perform computations locally at the AP, and serve only a subset of users \cite{Demir2021NOW,AMMAR2022COMST}.
Among these schemes, the maximum ratio transmission (MRT) technique, also known as conjugate beamforming \cite{NGO2017TWC} or coherent beamforming \cite{OZDOGAN2019TWC}, is popular in various system performance analyses due to its simple structure, ease of analysis, and availability of closed-form expressions \cite{HOANG2018TCOMM,PAPZAFEIROPOULOS2020TVT,ELHOUSHY2021COMST,MAI2022TWC,PARIDA2023TWC}. 
However, a significant drawback of MRT is its lack of consideration for inter-user interference suppression and the requirement for a very large number of antennas to achieve channel orthogonality, leading to poor performance in practical multi-user systems \cite{BJORNSON2020TWC,Demir2021NOW}.
On the other hand, the ZF scheme, due to its pseudo-inverse structure, cannot be directly computed locally at the AP and is not applicable when the number of users exceeds the number of antennas. 
Moreover, as the number of users approaches the number of antennas, the performance of ZF degrades sharply, limiting its application in cell-free systems \cite{Interdonato2020TWC}.
Conversely, local partial MMSE (LP-MMSE) precoding, also known as  local partial regularized zero-forcing (LP-RZF) precoding, has been widely used in research that does not focus on performance analysis due to its excellent scalability and performance
 \cite{AMMAR2022COMST,ZAHER2023TWC,DEMIR2024JSAC}.

However, the presence of a regularization term in LP-MMSE precoding makes it challenging to derive closed-form expressions for the performance of systems employing this precoding \cite{Interdonato2020TWC}.
This difficulty has hindered the investigation of resource allocation problems and performance analysis based on scalable cell-free systems for LP-MMSE precoding systems, leaving researchers to rely on the performance-limited MRT precoding \cite{HOANG2018TCOMM,PAPZAFEIROPOULOS2020TVT,ELHOUSHY2021COMST,MAI2022TWC,PARIDA2023TWC,AMMAR2022COMST}.
Fortunately, the development of large-dimensional random matrix theory has enabled the derivation of deterministic equivalents for the performance of MMSE precoding based system \cite{Bai2009WS,couillet2011CUP,Wagner2012TIT}. 
In existing works, large-dimensional random matrix theory has been employed to investigate the performance of various MMSE precoding based systems, including centralized systems with global information \cite{Wagner2012TIT}, distributed systems \cite{ZHANG2013TWC}, and other cell-free systems based on global cooperation, such as network-assisted full-duplex systems \cite{WANG2020TWC}, distributed channel information systems \cite{LI2020TIT}, asynchronous reception systems \cite{LI2021TVT}, uplink uncrewed aerial vehicle (UAV) networks \cite{Diaz2023TWC}, and uplink systems with non-ideal hardware \cite{XIE2024TWC}. 
However, research on the downlink of high-performance and scalable cell-free massive MIMO systems based on LP-MMSE remains unexplored.

Therefore, to address the challenges in applying the LP-MMSE scheme to the performance analysis of user-centric scalable cell-free massive MIMO systems, this paper leverages large-dimensional random matrix theory \cite{Bai2009WS,couillet2011CUP,Wagner2012TIT,ZHANG2013TWC} to derive the deterministic equivalent of the ergodic rate for user-centric scalable cell systems based on LP-MMSE precoding, as well as its approximate derivative with respect to the channel correlation matrix. 
Furthermore, based on the derivative, a deployment optimization algorithm is designed to enhance system performance by optimizing system deployment. 
The main contributions of this paper are summarized as follows:

\begin{itemize}
    \item Leveraging large-dimensional random matrix theory, this paper presents a universally applicable deterministic equivalent for the ergodic sum rate of a user-centric scalable cell-free massive MIMO system that considers arbitrary correlation Rayleigh channel, channel estimation errors, arbitrary power allocation, and an arbitrarily large number of antennas across multiple APs. 
    Numerical simulations demonstrate that this deterministic equivalent approximation achieves high accuracy even with 16 antennas per AP and gradually approaches complete convergence as the number of antennas increases.
    \item Furthermore, this paper derives the approximate derivative of the deterministic equivalent of the ergodic sum rate with respect to the channel correlation matrix, which includes large-scale information. This derivative enables a range of system optimizations related to channel correlation, such as AP location deployment, UAV  flight, array angle, and movable antenna optimization.
    \item Using above derivative, this paper designs a Barzilai-Borwein based gradient descent method to improve system deployment. Experimental results show that this algorithm can effectively enhance the sum rate performance by optimizing the deployment locations of APs.
\end{itemize}

{\bf{Notations}:}
$(\cdot)^\mathrm{T}$ and $(\cdot)^\mathrm{H}$refer to the transpose and Hermitian transpose respectively.
$\|\cdot\|_F$ denotes the 2-norm for vectors and the Frobenius norm for matrices.
$\mathcal{CN}(0,\sigma^2)$ indicates a complex Gaussian distribution with a zero mean and a variance of $\sigma^2$.
$\mathbf{I}_N$ is an $N \times N$ identity matrix.
$\mathbb{E}[\cdot]$ represent mathematical expectation.

\section{System model}
This paper considers a time division duplexing (TDD) cell-free MIMO system, where there are $K$ users and $M$ APs, with the $m$-th AP equipped with $N_m$ antennas, resulting in a total of $N=\sum_{m=1}^{M}N_m$ antennas in the system.
\begin{figure}[htbp]
    \centering
    \includegraphics[width=0.45\textwidth]{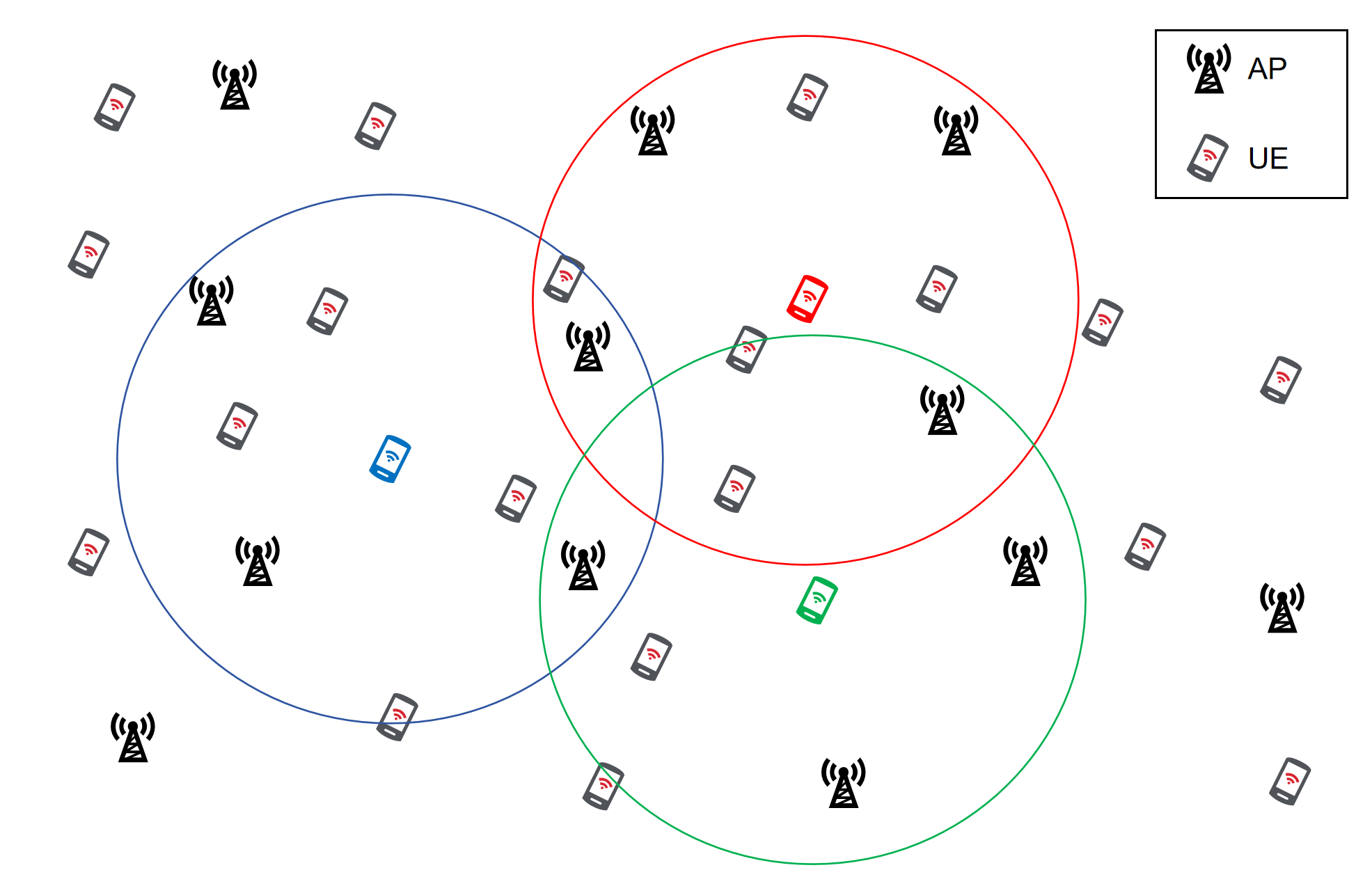}
    \caption{Cell-free system}\label{sys_model}
\end{figure}

As show in Fig.\ref{sys_model}, within a certain user-centric area, multiple APs jointly serve multiple users, and the large-scale information of user channels is determined by their locations and known nearby APs for further signal processing.
The users in the system utilize orthogonal pilots to estimate channels through the MMSE method, with a certain degree of channel estimation error.

\subsection{Channel model}
Assuming the channel between all users and all APs is denoted as $\mathbf{H} = \left [\mathbf{h}_{1},\mathbf{h}_{2},\cdots,\mathbf{h}_{K}  \right ]^{\mathrm{T}} \in \mathbb{C}^{K \times N}$, where $\mathbf{h}_k=[\mathbf{h}^{\mathrm{T}}_{k,1},\mathbf{h}^{\mathrm{T}}_{k,2},\cdots,\mathbf{h}^{\mathrm{T}}_{k,N}]^{\mathrm{T}}$ is the channel between user $k$ and all APs and $\mathbf{h}_{k,m}=[h_{k,m_1},h_{k,m_2},\cdots,h_{k,m_{N_m}}]^{\mathrm{T}} \in \mathbb{C}^{N_{m} \times 1}$ is the channel between user $k$ and AP $m$.
Assuming a Rayleigh channel, we have $\mathbf{h}_{k,m} \sim \mathcal{CN}(0,\mathbf{R}_{k,m})$, where $\mathbf{R}_{k,m} \in \mathbb{C}^{N_{m} \times N_{m}}$ is a positive semi-definite covariance matrix, and $\mathbf{R}_{k,m} = \beta_{k,l}\overline{\mathbf{R}}_{k,m}$ represents the correlation matrix between channels. 
Here, $\beta_{k,l}$ denotes the large-scale fading component determined by distance, and $\overline{\mathbf{R}}_{k,m}$ is the standard channel correlation matrix with diagonal elements equal to $1$.

More generally, it is assumed that the elements in $\mathbf{x}_{k,m} \in \mathbb{C}^{N_{m} \times 1}$ are independently and identically distributed elements following $\mathcal{CN}(0,\frac{1}{N_m} \mathbf{I}_{N_m})$.
Then, in this paper, we consider $\mathbf{h}_{k,m} = \mathbf{R}^{1/2}_{k,m}\mathbf{x}_{k,m}$.
Furthermore, the channels between APs located at different positions are unlikely to be correlated.
Therefore, the correlation matrix of the user's channel vectors is a block diagonal matrix which can be expressed as $\mathbf{h}_{k} = \mathrm{diag}{(\mathbf{R}^{1/2}_{k,1},\mathbf{R}^{1/2}_{k,2},\cdots,\mathbf{R}^{1/2}_{k,M})}\mathbf{x}_{k}$.

Although the channel is assumed to be Rayleigh here, by leveraging the conclusions about random phase Rician channels from \cite{Diaz2023TWC,Diaz2024TWC}, the estimated channel still follows a zero-mean Gaussian distribution. 
Therefore, the conclusions of this paper can be extended to Rician channel.

\subsection{Initial Access and Channel Estimation in User-Centric Scalable Systems}
The initial access method in this context can refer to the steps outlined in \cite{BJORNSON2020TWC}.
Initially, users measure the large-scale fading $\beta_{k,l}$ through synchronization signals periodically broadcast by surrounding APs. Simultaneously, they establish initial access with the AP with the strongest signal and synchronize the large-scale information of surrounding APs measured by the user \cite{Stefania2011Wiley,SANGUINETTI2012TWC}. 
This AP is referred to as the primary AP for the user.
Subsequently, the primary AP is responsible for managing services to the user, allocating pilots, and notifying a group of APs with stronger signals of the user to perform channel estimation.

This system adopts the MMSE channel estimation method utilizing orthogonal pilots. 
Assuming the length of the pilot sequence is $\tau_p$, we can design up to $\tau_p$ fully orthogonal pilot sequences. 
We denote the $l$-th pilot sequence as $\boldsymbol{\phi}_l \in \mathbb{C}^{\tau_p \times 1}$ with $||\boldsymbol{\phi}_l||^2_{F}=\tau_p$. Furthermore, based on the orthogonality assumption, we have
\begin{equation}\label{Phi}
    \boldsymbol{\phi}^{\mathrm{H}}_j \boldsymbol{\phi}_l=
    \begin{cases}
        \tau_p, & j=l          \\
        0,      & j \neq l   
    \end{cases}.
\end{equation}
In the system, users first transmit pilots to the AP to prepare for communication. 
The pilot signal received at $m$-th AP is given by $\mathbf{Y}^{\mathrm{p}}_{m}=\sum^{K}_{k=1} \sqrt{\eta_{k}} \mathbf{h}_{k,m} \boldsymbol{\phi}^{\mathrm{H}}_{t_{k}} + \mathbf{N}_{m}$.
Subsequently, using the orthogonality of the pilot sequences, interference between users utilizing different pilots is eliminated as
\begin{subequations}
    \begin{align}
        \mathbf{y}^{\mathrm{p}}_{k,m} & =\mathbf{Y}^{\mathrm{p}}_{m} \boldsymbol{\phi}_{t_{k}}/{\sqrt{\tau_{p}}}\\
         & =\sum^{K}_{i=1} \sqrt{\eta_{i}} \mathbf{h}_{i,m} \boldsymbol{\phi}^{\mathrm{H}}_{t_{i}} \boldsymbol{\phi}_{t_{k}} /{\sqrt{\tau_{p}}} + \mathbf{N}_{m} \boldsymbol{\phi}_{t_{k}} /{\sqrt{\tau_{p}}} \\
         & =\sqrt{\eta_{k}\tau_p}\mathbf{h}_{k,m}+\sum_{i \in \mathcal{P}_{k}\setminus {k}}\sqrt{\eta_{i}\tau_p}\mathbf{h}_{i,m}+\mathbf{n}_{t_k,m},
    \end{align}
\end{subequations}
where $\mathbf{n}_{t_k,m}=\mathbf{N}_m \boldsymbol{\phi}_{t_{k}}/{\sqrt{\tau_{p}}} \sim \mathcal{CN}(\mathbf{0}_N,{\sigma_{ul}^2}{\mathbf{I}_N})$, $t_k$ is the index of the pilot assigned to user $k$ and $\mathcal{P}_{k}=\{ i | t_i = t_k , i=1,\cdots,K \}$ is the user set that use the same pilot as user $k$. 
Using the MMSE channel estimation method, the channel estimation can be written as
$\widehat{\mathbf{h}}_{k,l}=\sqrt{\eta_{k}\tau_p}\mathbf{R}_{k,l}\boldsymbol{\Psi}^{-1}_{t_k,l} \mathbf{y}^{\mathrm{p}}_{k,m}$,
where$\boldsymbol{\Psi}_{t_k,l}=\mathbb{E}\{ \mathbf{y}^{\mathrm{p}}_{k,m} (\mathbf{y}^{\mathrm{p}}_{k,m})^{\mathrm{H}} \}=\sum_{i \in \mathcal{P}_{k}} \eta_{i} \tau_{p}\mathbf{R}_{i,l}+{\sigma_{ul}^2}{\mathbf{I}_N}$.

Subsequently, we can define the error channel as $\widetilde{\mathbf{h}}_{k,l}=\mathbf{h}_{k,l}-\widehat{\mathbf{h}}_{k,l}$, where
\begin{subequations}
    \begin{align}
        \widehat{\mathbf{h}}_{k,l} &= \widehat{\mathbf{R}}_{k,l}^{1/2}\widehat{\mathbf{x}}_{k,l} \sim \mathcal{CN}(\mathbf{0}_N,\widehat{\mathbf{R}}_{k,l})\\
        \widetilde{\mathbf{h}}_{k,l} &= \widetilde{\mathbf{R}}_{k,l}^{1/2}\widetilde{\mathbf{x}}_{k,l} \sim \mathcal{CN}(\mathbf{0}_N,\widetilde{\mathbf{R}}_{k,l}) 
    \end{align}
\end{subequations}
where $\widehat{\mathbf{R}}_{k,l} =\eta_k \tau_p \mathbf{R}_{k,l} \boldsymbol{\Psi}_{t_k,l} \mathbf{R}_{k,l}$ and $\widetilde{\mathbf{R}}_{k,l} =\mathbf{R}_{k,l}-\eta_k \tau_p \mathbf{R}_{k,l} \boldsymbol{\Psi}_{t_k,l} \mathbf{R}_{k,l}$.
Here, $\widetilde{\mathbf{x}}_{k,l}$ and $\widehat{\mathbf{x}}_{k,l}$ are independent and follow the same distribution $\mathcal{CN}(\mathbf{0}_N,\frac{1}{N_l} \mathbf{I}_{N_l})$. It is straightforward to infer that $\widehat{\mathbf{h}}_{k,l}$ and $\widetilde{\mathbf{h}}_{k,l}$ are independent \cite{Demir2021NOW}.

\subsection{Downlink Transmission Model}
Let the signal transmitted to the user be denoted as $\mathbf{q}=\left [q_1,q_2,\cdots,q_k \right ]^{\mathrm{T}}$ and assume that the transmitted signals from users are independent Gaussian random variables, $q_k \sim \mathcal{N}(0,1)$, thus $\mathbb{E}[q^2_k]=1$.

The precoding vector for user $k$ that includes power is denoted as $\mathbf{w}_k \in \mathcal{C}^{N \times 1}$.
Assuming the noise at the receiving end is $\mathbf{n}=\left [n_1,n_2,\cdots,n_k \right ]^{\mathrm{T}}$ and follows a zero-mean Gaussian distribution, the formula for the received signal vector by all users in the MIMO system is written as $\mathbf{y}=\mathbf{H}\mathbf{W}\mathbf{q}+\mathbf{n}$, where $\mathbf{y}=\left [y_1,y_2,\cdots,y_k \right ]^{\mathrm{T}}$, and $y_k$ represents the signal received by user $k$.
From the perspective of the received signal, the reception can be further expressed as
\begin{equation}
    y_k=\mathbf{h}^{\mathrm{H}}_k \mathbf{w}_k q_k+\sum^{K}_{i=1, i \neq k}\mathbf{h}^{\mathrm{H}}_k \mathbf{w}_i q_i + n_{k},
\end{equation}
where $\mathbf{h}^{\mathrm{H}}_k \mathbf{w}_k q_k$ represents the desired signal component after transmission, $\sum^{K}_{i=1, i \neq k}\mathbf{h}^{\mathrm{H}}_k \mathbf{w}_i q_i$ represents the inter-user interference, and $n_{k}$ represents the noise.

According to the Shannon capacity formula for the additive white Gaussian noise channel, the achievable rate for user $k$ can be expressed as $R_{k}=\log_2\left( 1 + \mathrm{SINR}_k\right)$, where 
\begin{equation}
    \mathrm{SINR}_k=\frac{|\mathbf{h}^{\mathrm{H}}_{k}\mathbf{w}_{k}|^2}{\sum_{i = 1, i \neq k}^K |\mathbf{h}^{\mathrm{H}}_{k}\mathbf{w}_{i}|^2+\sigma^{2}}.
\end{equation}
The ergodic sum rate of this system can be defined as:
\begin{equation}
    R_{\mathrm{sum}}=\sum_{k=1}^{K}\mathbb{E} \left[ \log_2 \left( 1 + \mathrm{SINR}_k \right) \right]
\end{equation}

\subsection{User-Centric Scalable LP-MMSE Transmission Method}
Assuming that the service set of AP $l$ is $\mathcal{S}_{l}$ and user $k$ is served by it, the corresponding precoding vector can be written as
\begin{equation}\label{LPMMSE}
    \mathbf{w}_{k,l}=\left(\sum_{k \in \mathcal{S}_l} \widehat{\mathbf{h}}_{k,l}\widehat{\mathbf{h}}_{k,l}^{\mathrm{H}}+\alpha \mathbf{I}_{N_{l}}\right)^{-1}\widehat{\mathbf{h}}_{k,l},
\end{equation}
where $\mathbf{w}_k=[\mathbf{w}^{\mathrm{T}}_{k,1},\mathbf{w}^{\mathrm{T}}_{k,2}, \cdots, \mathbf{w}^{\mathrm{T}}_{k,M}]^{\mathrm{T}}$ and $\alpha$ is a regularization  parameter. 
In an ideal global system, an appropriate value of $\alpha$ can optimize this linear precoding, but this optimal value often requires heuristic search. 
Alternatively, setting $\alpha=\sigma^2$ can directly achieve a good performance \cite{BJORNSON20142014MSP}.
Additionally, if user $k$ is not served by AP $l$, then $\mathbf{w}_{k,l}=\mathbf{0}$.

Let $\mathbf{W}_{l}=\left(\sum_{k \in \mathcal{S}_l} \widehat{\mathbf{h}}_{k,l}\widehat{\mathbf{h}}_{k,l}^{\mathrm{H}}+\alpha \mathbf{I}_{N_{l}}\right)^{-1}$. 
Under LP-MMSE precoding, the  signal-to-interference-plus-noise ratio (SINR) of user $k$ can be expressed as
\begin{equation}\label{SINR}
    \mathrm{SINR}_k=\frac{\left|\sum_{l=1}^{N} \mathbf{h}_{k,l}^H \mathbf{W}_{l} \widehat{\mathbf{h}}_{k,l}  \sqrt{\rho_{k,l} p_{k,l}}\right|^2}
    {\sum_{j = 1 ,j \neq k}^{K}\left|\sum_{l=1}^{N} \mathbf{h}_{k,l}^H \mathbf{W}_{l} \widehat{\mathbf{h}}_{j,l}  \sqrt{\rho_{j,l} p_{j,l}}\right|^2 + \sigma^2},
\end{equation}
where $\rho_{k,l}=1/\|\widehat{\mathbf{W}}_{l} \widehat{\mathbf{h}}_{k,l}\|_{\mathrm{F}}^2$ is the power normalization factor, and $p_{k,l}$ is the transmit power of AP $l$ to user $k$. 
When $p_{k,l}=0$, AP $l$ does not serve user $k$.

Furthermore, for convenient representation and derivation, we define a 0-1 variable $s_{k,l} \in \{0,1\}$, where $s_{k,l}=1$ indicates that user $k$ is served by AP $l$, and $s_{k,l}=0$ indicates that user $k$ is not served by AP $l$. 
Then set of all APs serving user $k$ is denoted as $\mathcal{B}_k=\{l|s_{k,l}=1\}$ and the set of user served by AP $l$ is denoted as $\mathcal{S}_l=\{k|s_{k,l}=1\}$.

Unless specified otherwise, the power allocation method in this system defaults to the following scalable approach:
\begin{equation}\label{pow_aloc}
    p_{k,l}=
    \begin{cases}
        P_{l} \frac{\sqrt{\beta_{k,l}}}{\sum_{j \in \mathcal{S}_{l}} \sqrt{\beta_{j,l}}} & \text{if } k \in \mathcal{S}_{l} \\
        0 & \text{otherwise}
    \end{cases},
\end{equation}
where $P_{l}$ is the transmit power constraint of AP $l$, and $\mathcal{S}_{l}$ is the set of users served by $l$. 
The specific power allocation scheme can also be obtained through optimization.

\section{Asymptotic Analysis of scalable cell-free massive MIMO systems with LP-MMSE Precoding}
In this section, we present the deterministic equivalent expression for SINR in scalable cell-free systems employing LP-MMSE precoding.

Firstly, we consider a large scale system where the number of antennas and users tend to infinity simultaneously, i.e., $\{\xi_i = N_i/K\}_{\forall i} \longrightarrow \infty$ and $0 < \lim \inf \xi_i \leq \lim \sup \xi_i < \infty$.
This assumption is denoted as $\mathcal{N} \rightarrow \infty$.
Subsequently, based on the large-dimensional random matrix theory, we can derive the deterministic equivalent of SINR as follows:

\begin{theorem}\label{main_theorem}
As $\mathcal{N} \rightarrow \infty$, the SINR of users in the system converges as follows
\begin{equation}
    \mathrm{SINR}_k -\overline{\mathrm{SINR}}_k
    \stackrel{N_l \rightarrow \infty}{\longrightarrow} 0
\end{equation}
almost surely, where
\begin{equation}
    \overline{\mathrm{SINR}}_k =
    \frac{\left |\sum_{l \in \mathcal{B}_k} \sqrt{p_{k,l}} \frac{e_{k,l}}{\sqrt{e'_{k,l}}} \right |^2}
    {\sum\limits_{l=1}^{N} \sum\limits_{j=1,j \neq k}^{K} p_{j,l} \frac{e'_{j,l,\widehat{\mathbf{R}}_{k,l}} + \left ( (1+e_{k,l})^2 e'_{j,l,\widetilde{\mathbf{R}}_{k,l}}\right )}
    {N_l e'_{j,l} (1+e_{k,l})^2}
    + \sigma^2}.
\end{equation}
The parameters $e_{k,l}$ and $\boldsymbol{\Psi}_{l}$ can be iteratively solved from the following fixed-point equations
\begin{subequations}
    \begin{align}\label{Psi_ekl}
        \boldsymbol{\Psi}_{l} & =\left ( \frac{1}{N_l}\sum_{i=1}^{K}\frac{s_{i,l}}{1+e_{i,l}}\widehat{\mathbf{R}}_{i,l}+\alpha\mathbf{I}_{N}\right )^{-1} \\
        e_{k,l}               & =\frac{s_{k,l}}{N_l}\tr(\widehat{\mathbf{R}}_{k,l}\boldsymbol{\Psi}_{l}).
    \end{align}
\end{subequations}

The vector parameter $e'_{k,l}$ can be directly solved from $\mathbf{e}'_{:,l}=\mathbf{J}_{l} \mathbf{e}'_{:,l}+\mathbf{v}_{l}$, where
\begin{subequations}\label{ejvl}
    \begin{align}
        \mathbf{e}'_{:,l}      & =[e'_{1,l},\cdots,e'_{K,l}]^{\mathrm{T}}\\
        [\mathbf{J}_{l}]_{i,j} & =\frac{s_{i,l}\tr{(\widehat{\mathbf{R}}_{i,l}\boldsymbol{\Psi}_{l} \widehat{\mathbf{R}}_{j,l}\boldsymbol{\Psi}_{l})}}{N_l^2(1+e_{j,l})^2}\\
        \mathbf{v}_{l}         & =\frac{1}{N_l}[\tr{(\widehat{\mathbf{R}}_{1,l}\boldsymbol{\Psi}_{l}\boldsymbol{\Psi}_{l})},\cdots,\tr{(\widehat{\mathbf{R}}_{K,l}\boldsymbol{\Psi}_{l} \boldsymbol{\Psi}_{l})}].
    \end{align}
\end{subequations}

The vector parameter $e'_{j,l,\widehat{\mathbf{R}}_{k,l}}$ can be directly solved from $\mathbf{e}'_{:,l,\widehat{\mathbf{R}}_{k,l}}=(\mathbf{I}_{N_l}-\mathbf{J}_{l})^{-1} \mathbf{v}_{k,l}$, where
\begin{subequations}\label{ejvkl}
    \begin{align}
        \mathbf{e}'_{:,l,\widehat{\mathbf{R}}_{k,l}} & =[e'_{1,l,\widehat{\mathbf{R}}_{k,l}},e'_{2,l,\widehat{\mathbf{R}}_{k,l}},\cdots,e'_{K,l,\widehat{\mathbf{R}}_{k,l}}]^{\mathrm{T}}\\
        [\mathbf{J}_{l}]_{j,i}                       & =\frac{s_{i,l}\tr{(\widehat{\mathbf{R}}_{j,l}\boldsymbol{\Psi}_{l} \widehat{\mathbf{R}}_{i,l}\boldsymbol{\Psi}_{l})}}{N_l^2(1+e_{i,l,\widehat{\mathbf{R}}_{k,l}})^2}\\
        \mathbf{v}_{k,l}                             & =\frac{1}{N_l}[\tr{(\widehat{\mathbf{R}}_{1,l}\boldsymbol{\Psi}_{l} \widehat{\mathbf{R}}_{k,l}\boldsymbol{\Psi}_{l})},\notag\\
        &\cdots,\tr{(\widehat{\mathbf{R}}_{K,l}\boldsymbol{\Psi}_{l} \widehat{\mathbf{R}}_{k,l}\boldsymbol{\Psi}_{l})}]^{\mathrm{T}}.
    \end{align}
\end{subequations}
\end{theorem}

The parameter $e'_{j,l,\widetilde{\mathbf{R}}_{k,l}} = \frac{1}{N_l} \tr{(\widetilde{\mathbf{R}}_{k,l} \partial\boldsymbol{\Psi}_{l,\widehat{\mathbf{R}}_{j,l}})}$ where
\begin{equation}
         \partial\boldsymbol{\Psi}_{l,\widehat{\mathbf{R}}_{j,l}} 
         =\boldsymbol{\Psi}_{l}
        \left( \sum_{i=1}^{K}\frac{e'_{i,l,\widetilde{\mathbf{R}}_{j,l}}\widehat{\mathbf{R}}_{i,l}}{N_l(1+e_{i,l})^2}+\widehat{\mathbf{R}}_{j,l} \right)
        \boldsymbol{\Psi}_{l}
\end{equation}

{\bf{Proof:}} See Appendix \ref{Prof_them_1}.

Furthermore, based on the definition of deterministic equivalents \cite{couillet2011CUP} and the continuous mapping theorem \cite{billingsley1995Wiley}, we have:
\begin{equation}
    \log_2 \left( 1 + \mathrm{SINR}_k \right) - \log_2 \left( 1 + \overline{\mathrm{SINR}}_k \right)
    \stackrel{\mathcal{N} \rightarrow \infty}{\longrightarrow} 0
\end{equation}
Let $\overline{R}_{\mathrm{sum}}=\sum_{k=1}^{K} \log_2 \left( 1 + \overline{\mathrm{SINR}}_k \right)$. Subsequently, we can obtain the deterministic equivalent of the ergodic sum-rate $R_{\mathrm{sum}}$ as:
\begin{equation}
    \frac{1}{K} \left( R_{\mathrm{sum}} - \overline{R}_{\mathrm{sum}} \right)
    \stackrel{\mathcal{N} \rightarrow \infty}{\longrightarrow} 0.
\end{equation}
Here, the ergodic sum rate $R_{\mathrm{sum}}$ is often obtained through Monte Carlo numerical simulations and lacks analytical tractability.
The determination of its deterministic equivalent $\overline{R}_{\mathrm{sum}}$ involves deterministic and explicit steps at each stage, facilitating system performance estimation, parameter solution, and optimization.

Moreover, unlike \cite{ZHANG2013TWC,WANG2020TWC,Diaz2023TWC}, in this paper, the power allocation variables are independent, allowing for flexible adoption of various power allocation schemes, such as the one based on large-scale information in \eqref{pow_aloc}, which is widely applicable to scalable systems employing different power allocation schemes.
\section{Deployment Optimization of User-Centric Scalable Cell-Free MIMO Systems}
This section presents the deterministic equivalent approximation of the derivative with respect to the channel correlation matrix in user-centric scalable cell-free MIMO systems. 
An optimization framework is established to enhance performance by adjusting the channel correlation matrix, which encompasses the system deployment including AP and user locations. 

\subsection{Problem Statement}

Given the position of AP $l$ as $\boldsymbol{\lambda}_l = (x_{l}^{AP}, y_{l}^{AP}) \in \mathbb{R}^{2 \times 1}$ and the position of user $k$ as $\boldsymbol{\lambda}_{\mathrm{U},k} = (x_{k}^{U}, y_{k}^{U})$, the channel correlation matrix is determined by their relative position and actual environment, denoted as $\mathbf{R}_{k,l} = \mathbf{R}_{k,l}(\boldsymbol{\lambda}_l)$. 
Once the channel correlation matrix is determined, the system performance can be analyzed using deterministic equivalents. 
Therefore, our optimization problem is formulated as follows

\begin{subequations}\label{MAXrate}
    \begin{align}
    \mathop{\max}_{\boldsymbol{\lambda}_1,\boldsymbol{\lambda}_2,\cdots,\boldsymbol{\lambda}_l} & \sum_{k=1}^{K}\mathbb{E} \left[ \log_2 \left( 1 + \mathrm{SINR}_k \right) \right].
    \end{align}
\end{subequations}

After the positions of APs change, it becomes difficult to directly obtain specific channel information at different locations. 
However, we can study the system performance at different locations using statistical channel modeling. 
Therefore, we adopt deterministic equivalents based on statistical channel information as the primary tool for performance analysis.
In this scenario, changes in the positions of users and APs primarily lead to variations in the channel correlation matrix which further affect the overall performance.

\subsection{Derivative of SINR with respect to the channel correlation matrix}
We need to obtain the expression for performance, specifically the variations of $\mathrm{SINR}_k$ with respect to $\mathbf{R}_{k,l}$ and $\mathbf{R}_{j,l}$ for $j \neq k$. 
To begin, we introduce some simplified notations
\begin{subequations}
    \begin{align}
        \mathrm{SA}_{k,l}    & =\frac{e_{k,l}}{\sqrt{e'_{k,l}}} \\
        \mathrm{ITF}_{k,j,l} & =\frac{e'_{k,l,\widehat{\mathbf{R}}_{j,l}} + (1+e_{k,l})^2 e'_{j,l,\widetilde{\mathbf{R}}_{k,l}}}
        {N_l e'_{j,l} (1+e_{k,l})^2} \\
        \mathrm{SINR}_k      & =\frac{(\sum_{l=1}^{N} \sqrt{p_{k,l}} \mathrm{SA}_{k,l})^2}{\sum_{j=1, j \neq k}^{K} \sum_{l=1}^{N} p_{j,l} \mathrm{ITF}_{k,j,l} + \sigma^2}.
    \end{align}
\end{subequations}

Since most parameters in the above equations contain the term $e_{k,l} = \frac{s_{k,l}}{N_l} \tr{(\widehat{\mathbf{R}}_{k,l}\boldsymbol{\Psi}_l)}$.
According to Eq. \eqref{Psi_ekl}, $\widehat{\mathbf{R}}_{k,l}$ items contained in $\boldsymbol{\Psi}_l$ will make the derivative solution involve complex multi-order implicit functions, resulting in the loss of practical value for precise derivative solutions.
In fact, many of these variables have minimal impact on the overall numerical value. 
Therefore, we will use the deterministic equivalent theory to process them.
Firstly, we take the derivative of the following equation for future use
\begin{align}
    &\frac{\partial \frac{1}{N_l}\tr{(\widehat{\mathbf{R}}_{k,l} (\sum_{j \in \mathcal{S}_l \setminus k} \widehat{\mathbf{R}}_{j,l}^{\frac{1}{2}}\mathbf{x}_{j,l}\mathbf{x}_{j,l}^{\mathrm{H}}\widehat{\mathbf{R}}_{j,l}^{\frac{1}{2}} + \alpha \mathbf{I}_{N_{l}})^{-1})}} {\partial \widehat{\mathbf{R}}_{k,l}} \notag\\
    &= \frac{1}{N_l} (\sum_{j \in \mathcal{S}_l \setminus k} \widehat{\mathbf{R}}_{j,l}^{\frac{1}{2}}\mathbf{x}_{j,l}\mathbf{x}_{j,l}^{\mathrm{H}}\widehat{\mathbf{R}}_{j,l}^{\frac{1}{2}} + \alpha \mathbf{I}_{N_{l}})^{-1}.
\end{align}

We revisit $\mathbf{W}_{l}=(\sum_{k \in \mathcal{S}_l} \widehat{\mathbf{h}}_{k,l}\widehat{\mathbf{h}}_{k,l}^{\mathrm{H}}+\alpha \mathbf{I}_{N_{l}})^{-1}$, and then define $\mathbf{A}_{[k],l}=\sum_{j=1, j \neq k}^{j \in \mathcal{S}_l}\widehat{\mathbf{h}}_{j,l}\widehat{\mathbf{h}}_{j,l}^{\mathrm{H}}+\alpha\mathbf{I}_{N}$.
According to Lemma \ref{B-Bvv}, we have
\begin{subequations}
    \begin{align}
         & \frac{1}{N_l}\tr{(\widehat{\mathbf{R}}_{k,l} \mathbf{A}_{[k],l}^{-1})}-\frac{1}{N_l}\tr{(\widehat{\mathbf{R}}_{k,l} \mathbf{W}_{l})}
        \stackrel{\mathcal{N} \rightarrow \infty}{\longrightarrow} 0 \\
         & {\Longleftrightarrow} \frac{\partial \frac{1}{N_l}\tr{(\widehat{\mathbf{R}}_{k,l} \mathbf{A}_{[k],l}^{-1})}} {\partial \widehat{\mathbf{R}}_{k,l}} -\frac{\partial \frac{1}{N_l}\tr{(\widehat{\mathbf{R}}_{k,l} \mathbf{W}_{l})}} {\partial \widehat{\mathbf{R}}_{k,l}}
        \stackrel{\mathcal{N} \rightarrow \infty}{\longrightarrow} 0 \\
        & {\Longleftrightarrow}\frac{1}{N_l} \mathbf{A}_{[k],l}^{-1}-\frac{\partial \frac{1}{N_l}\tr{(\widehat{\mathbf{R}}_{k,l} \boldsymbol{\Psi}_{l})}} {\partial \widehat{\mathbf{R}}_{k,l}}
        \stackrel{\mathcal{N} \rightarrow \infty}{\longrightarrow} 0\\
        & \Longleftrightarrow \frac{1}{N_l} \tr{((\mathbf{A}_{[k],l}^{-1}
        -\frac{\partial \tr{(\widehat{\mathbf{R}}_{k,l} \boldsymbol{\Psi}_{l})}} {\partial \widehat{\mathbf{R}}_{k,l}})\widehat{\mathbf{R}}_{k,l})}
        \stackrel{\mathcal{N} \rightarrow \infty}{\longrightarrow} 0
    \end{align}
\end{subequations}

Thus, we can obtain a key derivative as follows:
\begin{equation}
    \frac{1}{N_l} \tr{(\boldsymbol{\Psi}_l \widehat{\mathbf{R}}_{k,l})}
    -\frac{1}{N_l} \tr{(\frac{\partial \tr{(\widehat{\mathbf{R}}_{k,l} \boldsymbol{\Psi}_{l})}} {\partial \widehat{\mathbf{R}}_{k,l}} \widehat{\mathbf{R}}_{k,l})}
    \stackrel{\mathcal{N} \rightarrow \infty}{\longrightarrow} 0.
\end{equation}

Similarly, we can further generalize to obtain
\begin{subequations}
    \begin{align}
        &\frac{1}{N_l}\tr{(\widehat{\mathbf{R}}_{k,l}\partial\boldsymbol{\Psi}_{l})}\notag\\
        &- \frac{1}{N_l}\tr{(\widehat{\mathbf{R}}_{k,l} \frac{\partial\tr{(\widehat{\mathbf{R}}_{k,l}\partial\boldsymbol{\Psi}_{l})}} {\partial\widehat{\mathbf{R}}_{k,l}} )}
          \stackrel{\mathcal{N} \rightarrow \infty}{\longrightarrow} 0 \label{dRPsi}\\
        &\frac{1}{N_l}\tr{(\widehat{\mathbf{R}}_{k,l} \partial\boldsymbol{\Psi}_{l,\widehat{\mathbf{R}}_{j,l}})}\notag\\
        &- \frac{1}{N_l}\tr{(\widehat{\mathbf{R}}_{k,l} \frac{\partial\tr{(\widehat{\mathbf{R}}_{k,l}\partial\boldsymbol{\Psi}_{l,\widehat{\mathbf{R}}_{j,l}})}} {\partial\widehat{\mathbf{R}}_{k,l}} )}
          \stackrel{\mathcal{N} \rightarrow \infty}{\longrightarrow} 0 \label{dRklPsi}
    \end{align}
\end{subequations}

Furthermore, due to the more frequent utilization of the estimated channel correlation matrix than the real channel correlation matrix, we can initially compute the derivative with respect to the estimated channel correlation matrix $\widehat{\mathbf{R}}_{k,l}$ as Eq. \eqref{dR_est_dR}, where $\mathbf{R}_{j,l},j \neq k$. 
Subsequently we determine the derivative with respect to the real channel correlation $\mathbf{R}_{k,l}$ according to the chain rule for composite functions.
\begin{figure*}
    \begin{equation}\label{dR_est_dR}
        \frac{\partial\widehat{\mathbf{R}}_{k,l}}{\partial\mathbf{R}_{k,l}} = \eta_{k} \tau_{p}\mathbf{R}_{k,l}\left ( \sum_{i \in \mathcal{P}_{k}} \eta_{i} \tau_{p}\mathbf{R}_{i,l}+{\sigma_{ul}^2}{\mathbf{I}_N} \right )^{-1}
        \left (2\mathbf{I}_{N_l} - \eta_{k} \tau_{p}\mathbf{R}_{k,l}\left ( \sum_{i \in \mathcal{P}_{k}} \eta_{i} \tau_{p}\mathbf{R}_{i,l}+{\sigma_{ul}^2}{\mathbf{I}_N} \right )^{-1}\right )
    \end{equation}
    \noindent\rule[0.25\baselineskip]{\textwidth}{0.8pt}
\end{figure*}
And, it is derived that 
$
\frac{\partial\widetilde{\mathbf{R}}_{k,l}}{\partial\mathbf{R}_{k,l}} = \mathbf{I}_{N_l} - \frac{\partial\widehat{\mathbf{R}}_{k,l}}{\partial\mathbf{R}_{k,l}}
$.

So far, we have completed the preparatory work before deriving and can proceed to solve the derivatives.
First, $\mathrm{SA}_{k,l}$ represents the signal received by user $k$ from AP $l$, which is primarily influenced by the channel between user $k$ and AP-$l$, namely $\mathbf{R}_{k,l}$. 
The impact of any other single user on it is extremely minimal.
Using Eq. \eqref{dRPsi} and Eq. \eqref{dRklPsi}, we can obtain
\begin{equation}
        \frac{\partial \mathrm{SA}_{k,l}}{ \partial \mathbf{R}_{k,l}} -
        \frac{e'_{k,l}\boldsymbol{\Psi}_{l}-\frac{1}{2}e_{k,l} \partial\boldsymbol{\Psi}_{l}}{N_{l}(e'_{k,l})^{3/2}}
        \frac{\partial \widehat{\mathbf{R}}_{k,l}}{ \partial \mathbf{R}_{k,l}}
          \stackrel{\mathcal{N} \rightarrow \infty}{\longrightarrow} \mathbf{0} 
\end{equation}
and $\frac{\partial \mathrm{SA}_{k,l}}{ \partial \mathbf{R}_{j,l}}\stackrel{\mathcal{N} \rightarrow \infty}{\longrightarrow}\mathbf{0}$ where $\mathbf{R}_{j,l},j \neq k$.

Subsequently, we proceed to solve for the square of the interference signal strength $\mathrm{ITF}_{k,j,l}$ experienced by user $k$ due to the signal transmitted from AP $l$ aiming at user $j$. 
This quantity is primarily influenced by $\mathbf{R}_{k,l}$ and $\mathbf{R}_{j,l}$ shown in Eq. \eqref{dITF_Rkl} and Eq. \eqref{dITF_Rjl} both.
\begin{figure*}
    \begin{equation}\label{dITF_Rkl}
    \frac{\partial \mathrm{ITF}_{k,j,l}}{\partial \mathbf{R}_{k,l}}
    - \left (
    \frac{\partial\boldsymbol{\Psi}_{l,\widehat{\mathbf{R}}_{j,l}}(1+e_{k,l}) - 2 e'_{j,l,\widehat{\mathbf{R}}_{k,l}}\boldsymbol{\Psi}_{l}}{N_{l}^2 e'_{j,l} (1+e_{k,l})^3} \frac{\partial \widehat{\mathbf{R}}_{k,l}}{ \partial \mathbf{R}_{k,l}} +
    \frac{\partial\boldsymbol{\Psi}_{l,\widehat{\mathbf{R}}_{j,l}}} {N_l e'_{j,l}} \frac{\partial \widetilde{\mathbf{R}}_{k,l}}{ \partial \mathbf{R}_{k,l}} \right )
    \stackrel{\mathcal{N} \rightarrow \infty}{\longrightarrow} 0
\end{equation}
\noindent\rule[0.25\baselineskip]{\textwidth}{0.8pt}
\end{figure*}

\begin{figure*}
    \begin{equation}\label{dITF_Rjl}
    \frac{\partial \mathrm{ITF}_{k,j,l}}{\partial \mathbf{R}_{j,l}}-
    \left ( \frac{e'_{j,l} \partial\boldsymbol{\Psi}_{l,\widehat{\mathbf{R}}_{k,l}} - e'_{j,l,\widehat{\mathbf{R}}_{k,l}} \partial\boldsymbol{\Psi}_{l}}{ N_{l}^2 (1 + e_{k,l})^2 (e'_{j,l})^2 } +
    \frac{e'_{j,l} \partial\boldsymbol{\Psi}_{l,\widetilde{\mathbf{R}}_{k,l}} - e'_{j,l,\widetilde{\mathbf{R}}_{k,l}} \partial\boldsymbol{\Psi}_{l}}{ N_{l}^3 (e'_{j,l})^2 }
    \right ) \frac{\partial \widehat{\mathbf{R}}_{j,l}}{ \partial \mathbf{R}_{j,l}}\stackrel{\mathcal{N} \rightarrow \infty}{\longrightarrow} \mathbf{0}
\end{equation}
\noindent\rule[0.25\baselineskip]{\textwidth}{0.8pt}
\end{figure*}

It is noteworthy that when $\widehat{\mathbf{R}}_{j,l}$ undergoes minimal changes or only exhibits overall variations, such as when the updated value is in the form as $a \widehat{\mathbf{R}}_{j,l}, \forall a \in \mathbb{C}$, we have
\begin{subequations}
    \begin{align}
        \tr{\left( \left( e'_{j,l} \partial\boldsymbol{\Psi}_{l,\widehat{\mathbf{R}}_{k,l}} - e'_{j,l,\widehat{\mathbf{R}}_{k,l}} \partial\boldsymbol{\Psi}_{l} \right)\widehat{\mathbf{R}}_{j,l}\right)}      & = 0 \\
        \tr{\left( \left( e'_{j,l} \partial\boldsymbol{\Psi}_{l,\widetilde{\mathbf{R}}_{k,l}} - e'_{j,l,\widetilde{\mathbf{R}}_{k,l}} \partial\boldsymbol{\Psi}_{l} \right) \widehat{\mathbf{R}}_{j,l}\right)} & = 0.
    \end{align}
\end{subequations}
At this point, the impact of $\widehat{\mathbf{R}}_{j,l}$ on the system does not need to be considered.
Furthermore, the impact of $\mathbf{R}_{u,l}$, where $u \neq j,k$, on it is minimal. 
It can be derived that$ \frac{\partial \mathrm{ITF}_{k,j,l}}{ \partial \mathbf{R}_{u,l}}\stackrel{\mathcal{N} \rightarrow \infty}{\longrightarrow}\mathbf{0}$.
Therefore, we have the derivative of SINR with respect to the channel correlation matrix Eq. \eqref{dSINR_dRkl} 
\begin{figure*}
    \begin{equation}\label{dSINR_dRkl}
    \frac{\partial \mathrm{SINR}_k}{\partial \mathbf{R}_{k,l}}-
    \left(
    \frac{2(\sum\limits_{m=1}^{N} \sqrt{p_{k,m}} \mathrm{SA}_{k}) \sqrt{p_{k,l}} \frac{\partial \mathrm{SA}_{k,l}}{ \partial \mathbf{R}_{k,l}}}{\sum\limits_{j=1, j \neq k}^{K} \sum\limits_{m=1}^{N} p_{j,m} \mathrm{ITF}_{k,j,m} + \sigma^2}
    -\frac{(\sum\limits_{m=1}^{N} \sqrt{p_{k,m}} \mathrm{SA}_{k})^2 (\sum\limits_{j=1,j \neq k}^{K} p_{j,l}  \frac{\partial \mathrm{ITF}_{k,j,l}}{\partial \mathbf{R}_{k,l}})}{(\sum\limits_{j=1,j \neq k}^{K} \sum\limits_{m=1}^{N} p_{j,m} \mathrm{ITF}_{k,j,m} + \sigma^2)^2}
    \right)
    \stackrel{\mathcal{N} \rightarrow \infty}{\longrightarrow} 0
\end{equation}
\noindent\rule[0.25\baselineskip]{\textwidth}{0.8pt}
\end{figure*}
and
\begin{subequations}\label{dSINR_dRjl}
\begin{align}
    &\frac{\partial \mathrm{SINR}_k}{\partial \mathbf{R}_{j,l}}-\\
    &\frac{(\sum\limits_{m=1}^{N} \sqrt{p_{k,m}} \mathrm{SA}_{k})^2 (\sum\limits_{j=1,j \neq k}^{K} p_{j,l}  \frac{\partial \mathrm{ITF}_{k,j,l}}{\partial \mathbf{R}_{j,l}})}{(\sum\limits_{j=1,j \neq k}^{K} \sum\limits_{m=1}^{N} p_{j,m} \mathrm{ITF}_{k,j,m} + \sigma^2)^2}
    \stackrel{\mathcal{N} \rightarrow \infty}{\longrightarrow} 0.
\end{align}
\end{subequations}
So far, the variation of SINR with respect to the channel correlation matrix has been clarified. 
Various optimization methods utilizing first-order gradients, such as gradient descent methods and quasi-Newton methods, can be applied. 
For constrained optimization problems, penalty function methods and other approaches can be utilized for handling.

\subsection{Deployment Optimization Methods}

After clarifying the variations in SINR and the channel correlation matrix, it is necessary to establish the relationship between the channel correlation matrix and location. 
This specific relationship depends on the modeling of channel correlation. 
If the model includes deterministic parameters such as angles, spread, and antenna spacing, further optimization of these parameters can lead to improved system performance.
However, these aspects are not the primary focus of this paper. 
Instead, we will only consider a simple method for modeling the channel correlation matrix.

In this section, in order to address issues related to abnormal path loss values when distances approach zero and also account for differentiability with respect to distance \cite{Diaz2023TWC}, we make an assumption that the channel correlation matrix is solely a function of distance, i.e.,$\mathbf{R}_{k,m}=2^{\alpha}\overline{\beta}(1 + \|\boldsymbol{\lambda}_l-\boldsymbol{\lambda}_{\mathrm{U},k}\|_{\mathrm{F}}/d_0)^{-\alpha}\overline{\mathbf{R}} _{k,m}$,
where $d_0$ is the reference distance, and $\overline{\beta}$ is the path loss at the reference distance.
One obtains
\begin{equation}
    \frac{\partial \mathbf{R}_{k,m}} {\partial \boldsymbol{\lambda}_l}=
    - 2^{\alpha} \frac{\alpha \overline{\beta}}{d_0}  \overline{\mathbf{R}} _{k,m}  \left ( 1+\frac{d_{k,l}}{d_0} \right )^{- \alpha -1}  \partial d_{k,l},
\end{equation}
\begin{equation}
    \partial d_{k,l} = \tr \left( \frac{(\boldsymbol{\lambda}_l-\boldsymbol{\lambda}_{\mathrm{U},k})^{\mathrm{H}}}{\|\boldsymbol{\lambda}_l-\boldsymbol{\lambda}_{\mathrm{U},k} \|_{\mathrm{F}}}  \partial \boldsymbol{\lambda}_l\right).
\end{equation}

Setting $\overline{R}_k=\log_2 \left( 1 + \overline{\mathrm{SINR}}_k \right)$, we have
\begin{equation}\label{dR_dlambda}
    \frac{\partial \overline{R}_{\mathrm{sum}} }{\partial \boldsymbol{\lambda}_l} = \sum\limits_{k=1}^{K} \frac{\partial \overline{R}_k}{\partial \boldsymbol{\lambda}_l}=
    \sum\limits_{k=1}^{K}  \frac{(\ln{2})^{-1}\sum\limits_{j=1}^{K} \sum\limits_{m=1}^{M}\frac{\partial \mathrm{SINR}_k}{\partial \mathbf{R}_{j,m}}\frac{\partial \mathbf{R}_{j,m}}{\partial \boldsymbol{\lambda}_l}}
    {1 + \mathrm{SINR}_k} 
\end{equation}

Furthermore, the coordinates of the AP can be updated using the gradient descent method, which is expressed as $\boldsymbol{\lambda}_{l}^{t+1}=\boldsymbol{\lambda}_{l}^{t}-\delta^{t}\frac{\partial \overline{R}_{\mathrm{sum}}^{t} }{\partial \boldsymbol{\lambda}_{l}^{t}}$, where $\delta_{l}$ represents the step size, and $t$ denotes the $t$-th iteration. The step size $\delta_{l}$ can be obtained using the Barzilai-Borwein method in the next subsection, which is a quasi-Newton method.

\subsection{Barzilai-Borwein Gradient Descent Method}
The deployment of AP locations faces a severe challenge: users who are too close to a AP can lead to rapidly changing derivatives, where even slight changes in position can result in significant variations. Therefore, before iterative convergence, we choose to sequentially mask the derivatives with the largest and most drastic changes. Otherwise, the minor changes in APs far from users would be overwhelmed by these drastic changes.
Initially, to accelerate the speed of iterative convergence, we select the Barzilai-Borwein algorithm to set the step size \cite{Liu2021PKU}
\begin{equation}\label{deta_l_BB}
    \delta_{l}^{t}= \frac{(\boldsymbol{\lambda}_{l}^{t}-\boldsymbol{\lambda}_{l}^{t-1})^{\mathrm{H}} (\frac{\partial \overline{R}_{\mathrm{sum}}^{t} }{\partial \boldsymbol{\lambda}_{l}^{t}} - \frac{\partial \overline{R}_{\mathrm{sum}}^{t-1} }{\partial \boldsymbol{\lambda}_{l}^{t-1}})}
    {(\frac{\partial \overline{R}_{\mathrm{sum}}^{t} }{\partial \boldsymbol{\lambda}_{l}^{t}} - \frac{\partial \overline{R}_{\mathrm{sum}}^{t-1} }{\partial \boldsymbol{\lambda}_{l}^{t-1}})^{\mathrm{H}} (\frac{\partial \overline{R}_{\mathrm{sum}}^{t} }{\partial \boldsymbol{\lambda}_{l}^{t}} - \frac{\partial \overline{R}_{\mathrm{sum}}^{t-1} }{\partial \boldsymbol{\lambda}_{l}^{t-1}})}.
\end{equation}
Additionally, since the first-order derivative approximation is only valid within a certain domain, to prevent numerical anomalies caused by excessive step size variations, we impose a restriction. 
We limit the step size for each movement of the AP's position to be within the range of $[20,50]$ to ensure both the speed and accuracy of the optimization. 
Thus, we have$\delta_{M}^{t} = 50/\max{\|\frac{\partial \overline{R}_{\mathrm{sum}}^{t} }{\partial \boldsymbol{\lambda}_{l}^{t}}\|_{\mathrm{F}}}$ and $\delta_{m}^{t} = 20/\max{\|\frac{\partial \overline{R}_{\mathrm{sum}}^{t} }{\partial \boldsymbol{\lambda}_{l}^{t}}\|_{\mathrm{F}}}$.

The algorithm can be summarized as follows:

\begin{algorithm}[H]
    \caption{Barzilai-Borwein Gradient Descent Method (BBGD)}
    \label{alg:BBGD}
    \begin{algorithmic}[1]
        \State Initialization step
        \While{The result has not converged}
        \State Obtain the derivative $\frac{\partial \overline{R}_{\mathrm{sum}}^{t} }{\partial \boldsymbol{\lambda}_{l}^{t}}$ according to \eqref{dR_dlambda}
        \State Calculate the step size $\delta_{l}$ based on \eqref{deta_l_BB} and the upper and lower bounds $[\delta_{m}^{t},\delta_{M}^{t}]$
        \State Update the AP coordinates $\boldsymbol{\lambda}_{l}^{t}$ based on the step size $\delta_{l}$, and calculate the updated sum rate $\overline{R}_{\mathrm{sum}}^{t}(\boldsymbol{\lambda}_{l}^{t})$

        \While{$\overline{R}_{\mathrm{sum}}^{t}(\boldsymbol{\lambda}_{l}^{t}) \leq \overline{R}_{\mathrm{sum}}^{t-1}(\boldsymbol{\lambda}_{l}^{t-1})$}

        \If{$\delta_{l} \leq \delta_{m}^{t}$}
        \State Sequentially remove the derivative with the largest $\left\|\frac{\partial \overline{R}_{\mathrm{sum}}^{t} }{\partial \boldsymbol{\lambda}_{l}^{t}}\right\|_{\mathrm{F}}$
        \If{All derivatives are zeroed}
        \State Stop iteration
        \EndIf
        \EndIf
        \State Update $\delta_{l} \longleftarrow \frac{1}{2}\delta_{l}$ and normalize it to the bounds $[\delta_{m}^{t},\delta_{M}^{t}]$
        \EndWhile
        \EndWhile
    \end{algorithmic}
\end{algorithm}

\section{Simulations}
In this section, we use computer simulations to validate the accuracy of the  deterministic equivalent of ergodic sum-rate provided by Theorem \ref{main_theorem} in the preceding sections. 
We conduct tests with varying regularization coefficients, non-ideal CSI and varying scales of APs, users, and antenna counts to show the precision and applicability of this  deterministic equivalent.
Subsequently, we generate initial locations of APs using uniform distribution and clustering algorithms to verify the feasibility of our proposed deployment optimization algorithm. 

\subsection{Basic System Settings and Generation of Basic Parameters}
Unless specified otherwise, the default parameter settings for our system are listed in Table \ref{tab:default_parameters}.
These parameters will serve as a benchmark for the system, and when we focus on changes to certain parameters, the settings of the other parameters can be referenced from this table.

\begin{table}\label{tab:default_parameters}
\begin{center}
\caption{A Simple Table Example.}
\label{tab1}
\begin{tabular}{| c | c |}
\hline
Simulation radius $r$ & $1$ km\\
\hline
Background noise $\sigma^2$ & $94$ dBm\\
\hline 
Pilot power $\eta$ & $0.4$ W\\
\hline 
Pilot length $\tau_p$ & 10 \\
\hline 
Carrier frequency $f_c$ & 3000MHz \\
\hline 
Reference distance $d_0$ & 50m \\
\hline 
number of APs $M$ & 15 \\
\hline 
per UE associate & 10 \\
\hline 
per AP power $P_l$ & 10W \\
\hline 
per AP antennas $N_l$ & 32 \\
\hline 
number of UE $K$ & 40 \\
\hline 
\end{tabular}
\end{center}
\end{table}

We utilize the path loss model provided in \cite{ETSI2020TR12} to generate the path losses for the reference distance
\begin{equation}
    \mathrm{PL}(dB)=-35.4+34\log_{10}(d_0)+20\log_{10}(f_c).
\end{equation}
We refer to the algorithm in \cite{Bjornson2017NOW} to calculate the channel correlation matrix
\begin{subequations}
    \begin{align}
        &[\mathrm{R}_{k,l}]_{t,m}= \int_{+\infty}^{-\infty} e^{2 \pi d_{\mathrm{H}} (t-m) \sin(\theta + \delta)} \frac{1}{\sqrt{2 \pi} \sigma_{\theta}} e^{-\frac{\delta^2}{2\sigma_{\theta}^{2}}} d \delta \label{R_cau_o}
        \\
        &\approx e^{2 \pi j d_{\mathrm{H}}(t-m) \sin (\theta_{k,l})} 
        e^{-\frac{\sigma_{\theta}^{2}}{2} (2 \pi d_{\mathrm{H}} (t-m) \cos (\theta) )^2 } .\label{R_cau_fast}
    \end{align}
\end{subequations}
where $d_{\mathrm{H}}=1/2$ is the antenna spacing, the angular diffusion reference \cite{ETSI2020TR12} is set to $\sigma_{\theta}=0.3316$, and $\theta_{k,l}$ is the angle of arrival of user. 
It is calculated according to the relative position of the user and the AP.
Since the integral operation in Eq. \eqref{R_cau_o} is time-consuming, we choose the approximation algorithm of Eq. \eqref{R_cau_fast} to generate the channel correlation matrix. 

\subsection{Accuracy of Deterministic Equivalent Rate Analysis}
In Monte Carlo simulations, our channel correlation matrix and large-scale parameters are fixed, while the channel is randomly generated based on the channel correlation matrix. 
The locations of UEs and APs in the system are respectively assigned from two sets of random number pools pre-generated by uniform distribution in sequence. 
This is to ensure the consistency and comparability of different simulations.

Besides, the dashed line and "MC" represent the results of Monte Carlo numerical simulations, while the icons and "DE" represent the approximate results of deterministic equivalents.

\subsubsection{Impact of Number of Antennas}
In this section, we set up different antenna configurations to observe the impact of the numbers of users and antennas on the accuracy of asymptotic analysis.

\begin{figure}[htbp]
    \centering
    \includegraphics[width=0.45\textwidth]{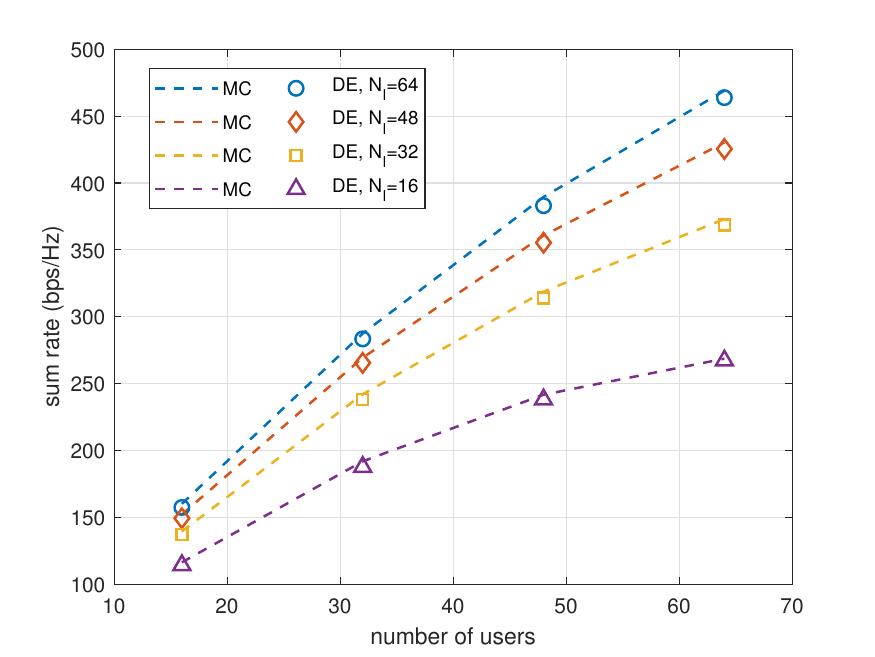}
    \caption{Impact of antenna number variation on system performance}\label{d_K_d_Ana_Cov_Est}
\end{figure}
Fig. \ref{d_K_d_Ana_Cov_Est} shows the sum rate as a function of the number of users for different numbers of antennas.
It can be seen from Fig. \ref{d_K_d_Ana_Cov_Est} that there is good approximation performance under various antenna and user configurations.
Since deterministic equivalent is an approximation obtained when the number of antennas approaches infinity, we usually have concerns about its performance when the number of antennas is relatively small. 
However, as can be seen from the Fig. \ref{d_K_d_Ana_Cov_Est}, our deterministic equivalent has a good approximation effect under a very practical antenna setting, ranging from 16 to 64 antennas.
Therefore, it can be concluded that the deterministic equivalent proposed in this paper is fully applicable to practical systems with imperfect CSI.

\subsubsection{Impact of AP Association Number}
In an ideal fully cooperative cell-free MIMO system, complete association and cooperation can achieve maximum performance, but they also bring exponentially increasing computational and backhaul demands.
In our user-centric scalable system, even without CSI sharing between APs and thus without achieving cooperative beamforming, good performance improvements can still be obtained through LP-MMSE precoding and cooperative transmission alone.

\begin{figure}[htbp]
    \centering
    \includegraphics[width=0.45\textwidth]{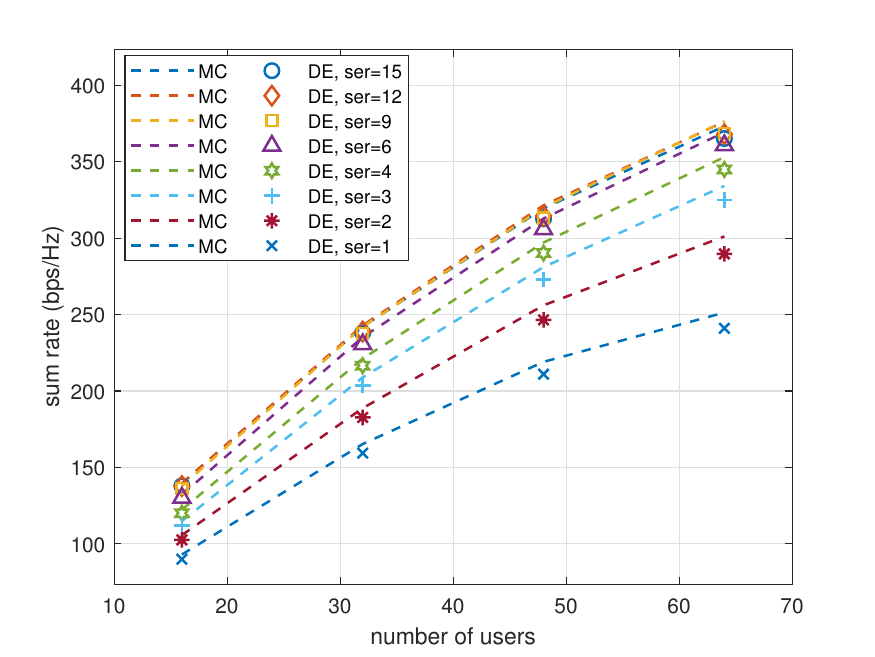}
    \caption{Impact of association number variation on system performance}\label{dK_dser_Cov_Est}
\end{figure}

As can be seen from Fig. \ref{dK_dser_Cov_Est}, this gain increases with the number of associations, but when the association number is around 6, the performance gain tends to be saturated. 
This may be due to non-cooperative beamforming and channel estimation errors. Therefore, when considering practical scalable systems, a limited scope and number of cooperations can themselves achieve the optimal state of such a system, without the need for comprehensive cooperation.

\subsubsection{Impact of regularization parameter}
\begin{figure}[htbp]
    \centering
    \includegraphics[width=0.45\textwidth]{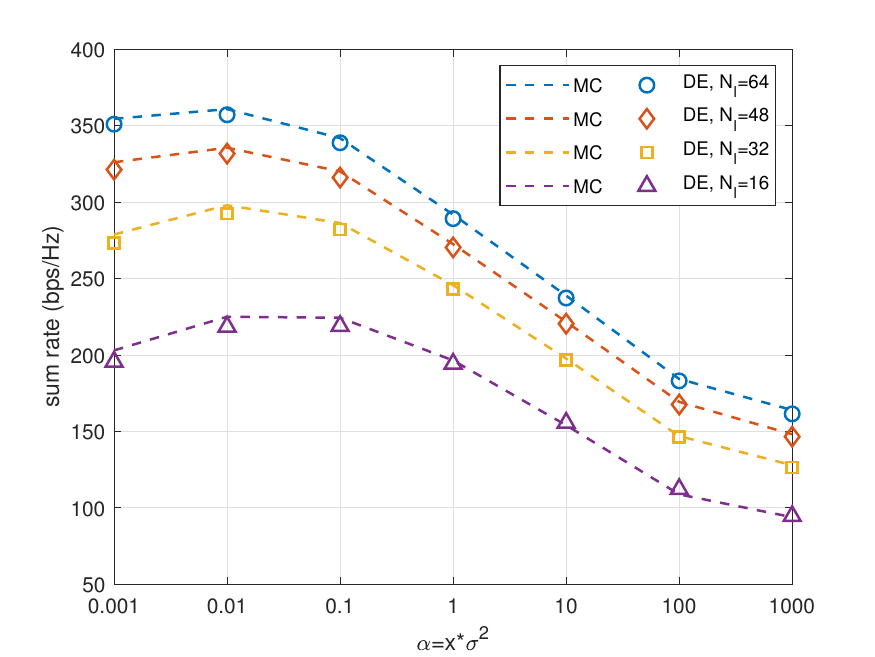}
    \caption{Impact of regularization parameter variation on system performance}\label{da_dana_Cov_Est}
\end{figure}

We investigate the impact of regularization parameter on the accuracy of the asymptotic analysis in ideal global MIMO systems, where RZF is proven to have the optimal precoding structure. 
By selecting appropriate regularization parameters, the optimal precoding design can be obtained. 
In the scalable system with local LP-MMSE, this structure still has strong performance.
In previous studies, the optimal regularization parameter can also be obtained through optimization methods using the deterministic equivalent analytical results. Since this is not the focus of this paper, we only verify the accuracy of the deterministic equivalent under different regularization parameter selections.
In this simulation, the RZF regularization parameter $\alpha$ is setting as $\alpha = x \sigma^2$ and $x$ is the x-axis in Fig. \ref{da_dana_Cov_Est}.

As can be seen from the Fig. \ref{da_dana_Cov_Est}, the deterministic equivalent approximation proposed in this paper has good approximation effects under different regularization parameter $\alpha$ selections. 
In fact, due to the structural characteristics of LP-MMSE precoding in Eq. \ref{LPMMSE}, as $\alpha \to 0$, RZF precoding tends to ZF precoding, while as $\alpha \to \infty$, RZF precoding tends to MRT precoding. 
This characteristic can also be used to further extend the conclusions to systems using MRT or ZF precoding.

\subsection{Deployment Optimization of User-Centric Scalable CF System Based on Progressive Analysis}
To verify the potential performance gains that the optimized deployment strategy may bring, this study utilizes randomly generated potential user points as representatives of users.
The settings of these user points can be flexibly adjusted according to the actual user demands in different areas to simulate the characteristics of user distribution in practical scenarios.
For instance, in areas such as bus stops, a higher density of user points may be set to simulate situations where people gather and stay; while in areas like sidewalks and parks, a lower density of user points is set to reflect the scattered distribution of people during most periods.

Due to the large system scale selected for performance analysis in the previous context, which makes it difficult to observe changes in location, in the deployment optimization, we choose a smaller system for observation. Specifically, we set the radius of the small cell to $r=300\mathrm{m}$, the number of AP antennas to $N_l=32$, the number of APs to $M=10$, the association number to $ser=6$, and the number of users to $K=48$. All user locations are randomly generated again, and the initial locations of APs are either uniformly distributed or based on a clustering algorithm, with other parameters remaining unchanged.
\subsubsection{Random Initial Positions}

\begin{figure*}[htbp]
\centering
\subfloat[Diagram of AP location optimization]{\includegraphics[width=0.5\textwidth]{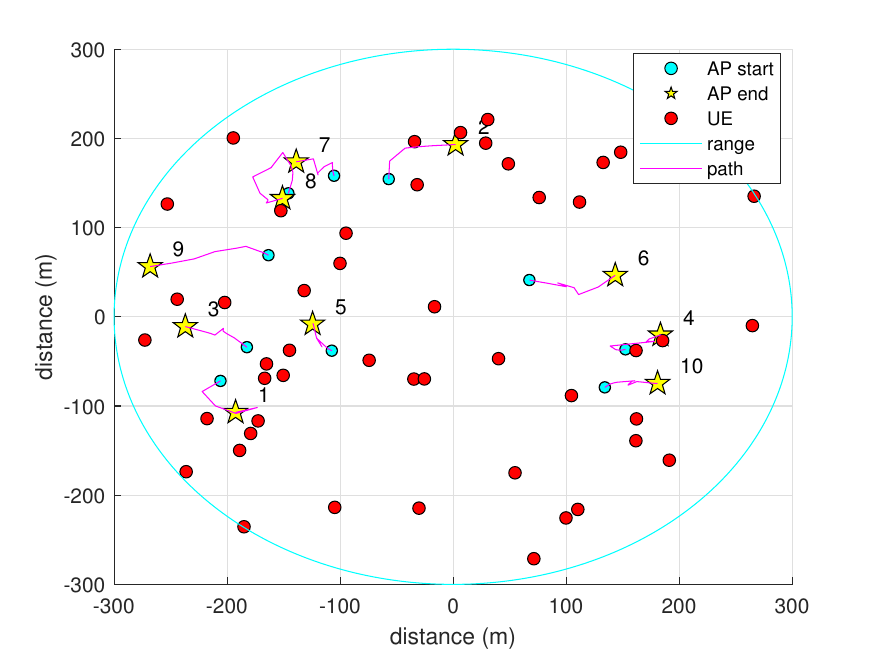}%
\label{bs_opt_base_random}}
\subfloat[Iterative graph of AP location optimization]{\includegraphics[width=0.5\textwidth]{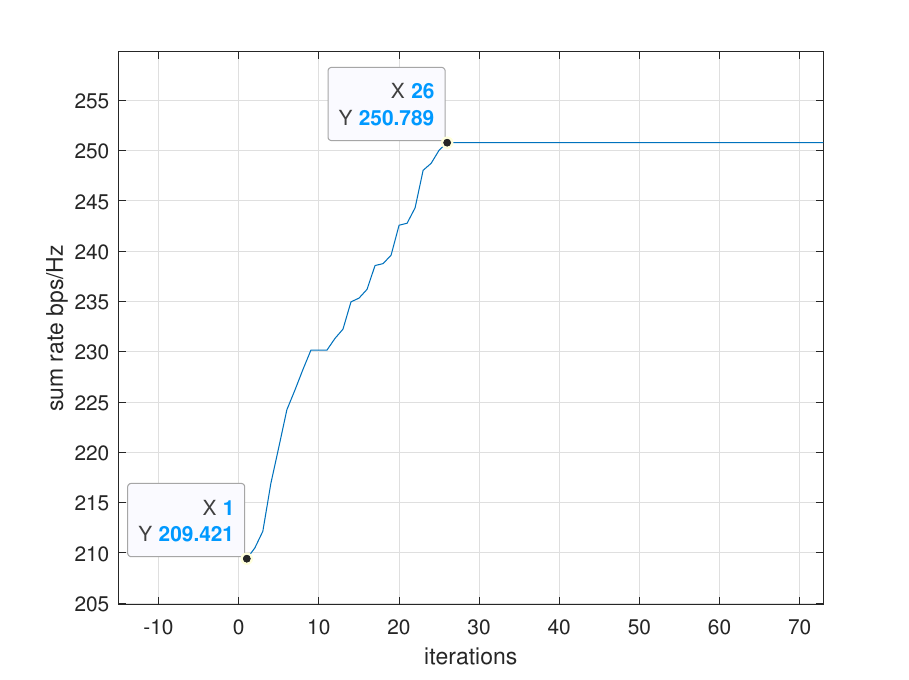}%
\label{bs_opt_base_random_ite}}
\caption{Optimization of initial random position}
\end{figure*}
We first observe the scenario with random initial positions to evaluate the performance of our algorithm in a more general context.
In Fig. \ref{bs_opt_base_random}, blue dots represent the initial positions of APs, yellow pentagrams indicate the final optimized positions of APs, red dots denote user positions, and the magenta lines connecting them represent the optimization paths.
As can be seen from Fig. \ref{bs_opt_base_random_ite}, our algorithm achieves good optimization results in the case of random initial positions, improving system performance from 209.4 bps/Hz to 250.8 bps/Hz within 26 iterations.
Since the algorithm operates based on the approximation of first-order derivatives, we can almost surely confirm that it reaches a local optimum.

Specifically, almost all APs undergo some degree of position optimization. 
Notably, AP1 and AP2 significantly move closer to densely populated user areas that lack APs, while AP4, 6 and 10 are optimized towards the edge regions to enhance signal coverage. 
Interestingly, AP8 appears to move in a circular trajectory and returns to its original position, which may be influenced by the movement of AP7.
In summary, our algorithm can effectively improve system performance by optimizing the deployment positions of APs. 
However, it is also evident that due to the complexity of the signal patterns in cell-free systems, the optimization paths can be relatively complex and non-intuitive.

\subsubsection{Cluster Initial Positions}
In the K-means clustering algorithm, the algorithm iteratively optimizes the centroid position of each cluster until a certain convergence condition is met, such as the change in centroid position being less than a certain threshold or reaching the maximum number of iterations. 
This is also the foundation of most AP deployment algorithms. 
In this section, we use the K-means clustering algorithm to cluster user points and consider each cluster center as a AP deployment location. 
This serves as our starting position and performance baseline to observe the optimization effect of our algorithm in this scenario.

\begin{figure*}[htbp]
\centering
\subfloat[Diagram of AP location optimization]{\includegraphics[width=0.5\textwidth]{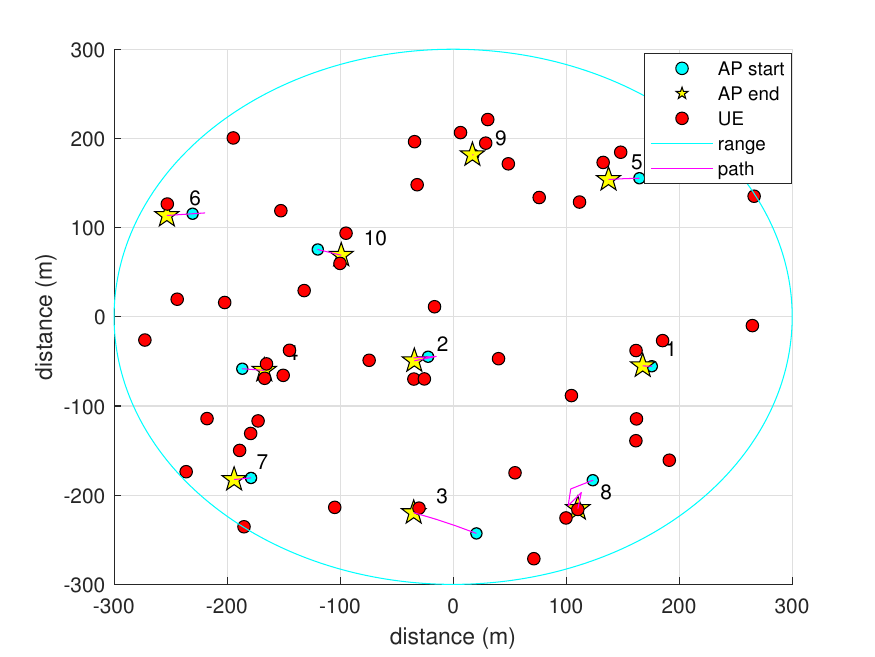}%
\label{bs_opt_base_kmeans}}
\subfloat[Iterative graph of AP location optimization]{\includegraphics[width=0.5\textwidth]{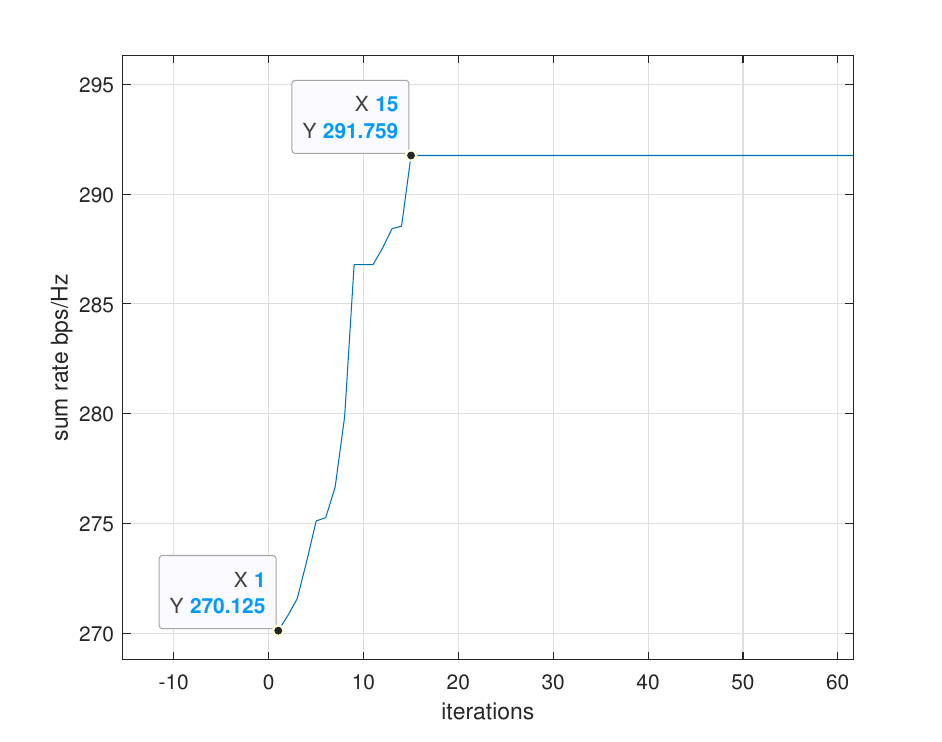}%
\label{bs_opt_base_kmeans_ite}}
\caption{Optimization of initial clustering position}
\end{figure*}

As can be seen from the iteration diagram Fig. \ref{bs_opt_base_kmeans_ite}, after about 60 iterations, the system performance is improved from 270.1 bps/Hz to 291.8 bps/Hz and reaches equilibrium.
From Fig. \ref{bs_opt_base_kmeans}, it can be seen that compared to the scenario with random initial positions, the optimization amplitude is generally not significant under the clustering initial positions. 
This also indirectly demonstrates the stability and efficiency of the classic K-means clustering algorithm in AP deployment optimization. 
However, our algorithm still effectively improves the system performance under this relatively optimal layout, which illustrates the universality of our algorithm.
Since our algorithm is designed based on derivatives, it has great optimization potential in any system that has not been proven to reach a local optimum. 
Therefore, our algorithm can be applied to the subsequent adjustment and further optimization of most optimization methods, enabling them to reach locally optimal positions.

\section{Conclusion}
This paper focused on the performance and optimization of user-centric scalable cell-free massive MIMO systems, considering correlated Rayleigh fading channels, imperfect CSI, and the LP-MMSE precoding method. 
Leveraging large-dimensional random matrix theory, we derived the deterministic equivalent of the ergodic sum-rate and its derivative with respect to the channel correlation matrix, providing theoretical support for system performance analysis and optimization. 
Based on this, we designed a gradient descent method inspired by the Barzilai-Borwein algorithm. 
By optimizing the deployment positions of access points, the ergodic sum-rate of the system was effectively enhanced. 
Simulation results demonstrated that the deterministic equivalent expressions are quite accurate, and our deployment optimization algorithm can effectively improve the performance of systems with different initial layouts.

{\appendix}\label{AppendixA}
\subsection{Proof of Theorem 1}\label{Prof_them_1}
For the cell-free system employing local partial MMSE precoding, further extension can be made based on Theorem \ref{Psi_e_k} from \cite{Wagner2012TIT} and \cite{ZHANG2013TWC} as
\begin{subequations}\label{LP_Psi_e_k} 
    \begin{align}
         & \boldsymbol{\Psi}_{l}=\left ( \frac{1}{N_l}\sum_{i=1}^{K}\frac{s_{i,l}}{1+e_{i,l}}\widehat{\mathbf{R}}_{i,l}+\alpha\mathbf{I}_{N}\right )^{-1} \\
         & e_{k,l}=\frac{s_{k,l}}{N_l}\tr(\widehat{\mathbf{R}}_{k,l}\boldsymbol{\Psi}_{l}). \label{ekl_th1}
    \end{align}
\end{subequations}

 As $\mathcal{N} \rightarrow \infty$, we have
\begin{equation}\label{QBaIQPsi}
    \frac{1}{N}\tr{(\mathbf{Q}(\mathbf{B}_l+\alpha\mathbf{I}_{N})^{-1})}
    - \frac{1}{N}\tr{(\mathbf{Q}\boldsymbol{\Psi}_{l})}
        \stackrel{\mathcal{N} \rightarrow \infty}{\longrightarrow} 0,
\end{equation}
almost surely, where $\mathbf{Q}$ is arbitrary matrix with uniformly bounded spectral norm and 
$\mathbf{B}_l=\sum_{k \in \mathcal{S}_l} \widehat{\mathbf{R}}_{k,l}^{\frac{1}{2}}\mathbf{x}_{k,l}\mathbf{x}_{k,l}^{\mathrm{H}}\widehat{\mathbf{R}}_{k,l}^{\frac{1}{2}}$.

For ease of reading, expressions such as $A \stackrel{N \rightarrow  \infty}{\longrightarrow} B$ will be used in the derivations below, as an alternative to the more rigorous notation $A-B \stackrel{N \rightarrow  \infty}{\longrightarrow} 0$.

\subsection*{Deterministic Equivalent of Signals}
In this section, we solve for the  deterministic equivalent expression of the signal amplitude $\mathbf{h}_{k,l}^H \mathbf{W}_{l} \widehat{\mathbf{h}}_{k,l}$. 
For simplicity, we initially assume $s_{k,l}=1$.
Let $\mathbf{A}_{[k],l}=\sum_{j=1, j \neq k}^{j \in \mathcal{S}_l} \widehat{\mathbf{h}}_{j,l}\widehat{\mathbf{h}}_{j,l}^{\mathrm{H}}+\alpha\mathbf{I}_{N}$. 
Then, we have $\mathbf{W}_{l}=\left ( \mathbf{A}_{[k],l}+\widehat{\mathbf{h}}_{k,l}\widehat{\mathbf{h}}_{k,l}^{\mathrm{H}} \right )^{-1}$. 
Subsequently, with Lemma \ref{xAx}, \ref{xAy} and \ref{qAq} we obtain
\begin{subequations}
    \begin{align}
         & (\mathbf{h}_{k,l}^{\mathrm{H}} )\mathbf{W}_{l}\widehat{\mathbf{h}}_{k,l}\\
         & =\frac{(\widehat{\mathbf{x}}_{k,l}^{\mathrm{H}}\widehat{\mathbf{R}}^{1/2}_{k,l} +\widetilde{\mathbf{x}}_{k,l}^{\mathrm{H}}\widetilde{\mathbf{R}}^{1/2}_{k,l}) \mathbf{A}_{[k],l}^{-1}\widehat{\mathbf{R}}^{1/2}_{k,l}\widehat{\mathbf{x}}_{k,l}}{1+\widehat{\mathbf{x}}_{k,l}^{\mathrm{H}}\widehat{\mathbf{R}}^{1/2}_{k,l}\mathbf{A}_{[k],l}^{-1}\widehat{\mathbf{R}}^{1/2}_{k,l}\widehat{\mathbf{x}}_{k,l}} \\
         & \stackrel{\mathcal{N} \rightarrow \infty}{\longrightarrow} \frac{(\widehat{\mathbf{x}}_{k,l}^{\mathrm{H}}\widehat{\mathbf{R}}^{1/2}_{k,l}) \mathbf{A}_{[k],l}^{-1}\widehat{\mathbf{R}}^{1/2}_{k,l}\widehat{\mathbf{x}}_{k,l}}{1+\widehat{\mathbf{x}}_{k,l}^{\mathrm{H}}\widehat{\mathbf{R}}^{1/2}_{k,l}\mathbf{A}_{[k],l}^{-1}\widehat{\mathbf{R}}^{1/2}_{k,l}\widehat{\mathbf{x}}_{k,l}}                           \\
         & \stackrel{\mathcal{N} \rightarrow \infty}{\longrightarrow}\frac{\frac{1}{N_l}\tr(\widehat{\mathbf{R}}_{k,l}\mathbf{A}_{[k],l}^{-1})}{1+\frac{1}{N}\tr(\widehat{\mathbf{R}}_{k,l}\mathbf{A}_{[k],l}^{-1})}\\
         & \stackrel{\mathcal{N} \rightarrow \infty}{\longrightarrow} \frac{\frac{1}{N_l}\tr(\widehat{\mathbf{R}}_{k,l}\boldsymbol{\Psi}_{l})}{1+\frac{1}{N_l}\tr(\widehat{\mathbf{R}}_{k,l}\boldsymbol{\Psi}_{l})}
    \end{align}
\end{subequations}
Therefore, the  deterministic equivalent expression of the signal is
\begin{equation}
    \mathbf{h}_{k,l}^{\mathrm{H}} \mathbf{W}_{l}\mathbf{h}_{k,l}
    \stackrel{\mathcal{N} \rightarrow \infty}{\longrightarrow} \frac{\frac{1}{N_l}\tr(\widehat{\mathbf{R}}_{k,l}\boldsymbol{\Psi}_{l})}{1+\frac{1}{N_l}\tr(\widehat{\mathbf{R}}_{k,l}\boldsymbol{\Psi}_{l})}=\frac{e_{k,l}}{1+e_{k,l}}
\end{equation}
In the Eq. \eqref{ekl_th1}, $e_{k,l}$ includes the case where $s_{k,l}=0$, thus it is applicable to all associated scenarios.
\subsection*{Deterministic Equivalent of Interference}
This section derives the  deterministic equivalent expression for $\sum_{j=1, j \neq k}^{K}|\sum_{l=1}^{N} \mathbf{h}_{k,l}^{\mathrm{H}} \mathbf{W}_{l} \mathbf{h}_{j,l}  \sqrt{\rho_{j,l} p_{j,l}}|^2$.
First, we have
\begin{subequations}
    \begin{align}
         & \sum_{j=1, j \neq k}^{K}|\sum_{l=1}^{N} \mathbf{h}_{k,l}^{\mathrm{H}} \mathbf{W}_{l} \mathbf{h}_{j,l}  \sqrt{\rho_{j,l} p_{j,l}}|^2          \\
         & =\sum_{j=1, j \neq k}^{K}(\sum_{l=1}^{N}\sum_{g=1}^{N}\sqrt{\rho_{j,l} p_{j,l}\rho_{j,g} p_{j,g}}  \mathbf{h}_{k,l}^{\mathrm{H}} \mathbf{W}_{l} \widehat{\mathbf{h}}_{j,l}   \widehat{\mathbf{h}}_{j,g}^{\mathrm{H}} \mathbf{W}_{g} \mathbf{h}_{k,g})
    \end{align}
\end{subequations}
We will derive the expressions by considering several different association relationships between users and  APs.
\subsubsection{Non-coherent Terms of Interference}
First, for the part where $l \neq g$, with Lemma \ref{xAx} and \ref{qAq} we have
\begin{subequations}
    \begin{align}
         & \mathbf{h}_{k,l}^{\mathrm{H}} \mathbf{W}_{l} \widehat{\mathbf{h}}_{j,l}   \widehat{\mathbf{h}}_{j,g}^{\mathrm{H}} \mathbf{W}_{g} \mathbf{h}_{k,g}\\
         & = \widehat{\mathbf{h}}_{j,g}^{\mathrm{H}} \mathbf{W}_{g} \mathbf{h}_{k,g} \mathbf{h}_{k,l}^{\mathrm{H}} \mathbf{W}_{l} \widehat{\mathbf{h}}_{j,l}\\
         & =\frac{\widehat{\mathbf{x}}_{j,g}^{\mathrm{H}} \left (\widehat{\mathbf{R}}_{j,g}^{\frac{1}{2}} \mathbf{A}_{[j],g}^{-1} \mathbf{h}_{k,g} \mathbf{h}_{k,l}^{\mathrm{H}} \mathbf{A}_{[j],l}^{-1} \widehat{\mathbf{R}}_{j,l}^{\frac{1}{2}} \right ) \widehat{\mathbf{x}}_{j,l}}
        {\left ( 1+\widehat{\mathbf{h}}_{j,g}^{\mathrm{H}}\mathbf{A}_{[j],g}^{-1}\widehat{\mathbf{h}}_{j,g} \right )
        \left ( \widehat{\mathbf{h}}_{j,l}^{\mathrm{H}} \mathbf{A}_{[j],l}^{-1} \widehat{\mathbf{h}}_{j,l}+1 \right )} \\
         & \stackrel{\mathcal{N} \rightarrow \infty}{\longrightarrow}
        \frac{\widehat{\mathbf{x}}_{j,g}^{\mathrm{H}} \left (\widehat{\mathbf{R}}_{j,g}^{\frac{1}{2}} \mathbf{A}_{[j],g}^{-1} \mathbf{h}_{k,g} \mathbf{h}_{k,l}^{\mathrm{H}} \mathbf{A}_{[j],l}^{-1} \widehat{\mathbf{R}}_{j,l}^{\frac{1}{2}} \right ) \widehat{\mathbf{x}}_{j,l}}
        {(1+e_{j,g})(1+e_{j,l})}
    \end{align}
\end{subequations}
In the above equation, the deterministic equivalents of the two parts in the denominator are both finite real numbers. Meanwhile, $\widehat{\mathbf{x}}_{j,g}^{\mathrm{H}}$ and $\widehat{\mathbf{x}}_{j,l}$ in the numerator are independent, and both of them are also independent of all elements in $\left (\widehat{\mathbf{R}}_{j,g}^{\frac{1}{2}} \mathbf{A}_{[j],g}^{-1} \mathbf{h}_{k,g} \mathbf{h}_{k,l}^{\mathrm{H}} \mathbf{A}_{[j],l}^{-1} \widehat{\mathbf{R}}_{j,l}^{\frac{1}{2}} \right )$. According to Lemma \ref{xAy}, the deterministic equivalent of the numerator, as well as the overall expression, is $0$.
\subsubsection{Coherent Terms of Interference}
For the part where $l=g$ and $k \in \mathcal{S}_l$, with Eq. \eqref{QBaIQPsi}, Lemma \ref{xAx} and \ref{qAq} we have
\begin{subequations}\label{hwhhwh}
    \begin{align}
         & \mathbf{h}_{k,l}^{\mathrm{H}} \mathbf{W}_{l} \widehat{\mathbf{h}}_{j,l}   \widehat{\mathbf{h}}_{j,l}^{\mathrm{H}} \mathbf{W}_{l} \mathbf{h}_{k,l} = \widehat{\mathbf{h}}_{j,l}^{\mathrm{H}} \mathbf{W}_{l} \mathbf{h}_{k,l} \mathbf{h}_{k,l}^{\mathrm{H}} \mathbf{W}_{l} \widehat{\mathbf{h}}_{j,l} \\
         & = \frac{ \widehat{\mathbf{x}}_{j,l}^{\mathrm{H}} \widehat{\mathbf{R}}^{1/2}_{j,l} \mathbf{A}_{[j],l}^{-1} \mathbf{h}_{k,l} \mathbf{h}_{k,l}^{\mathrm{H}} \mathbf{A}_{[j],l}^{-1}  \widehat{\mathbf{R}}^{1/2}_{j,l} \widehat{\mathbf{x}}_{j,l}}
        {\left (1+\widehat{\mathbf{x}}_{j,l}^{\mathrm{H}} \widehat{\mathbf{R}}^{1/2}_{j,l} \mathbf{A}_{[j],l}^{-1} \widehat{\mathbf{R}}^{1/2}_{j,l} \widehat{\mathbf{x}}_{j,l} \right )^2} \\
         & \stackrel{\mathcal{N} \rightarrow \infty}{\longrightarrow}
        \frac{1}{\left (1+e_{j,l} \right )^2} \frac{1}{N_l} \tr{(\widehat{\mathbf{R}}_{j,l} \mathbf{W}_l \mathbf{h}_{k,l} \mathbf{h}_{k,l}^{\mathrm{H}} \mathbf{W}_l)} \\
         & = \frac{\tr{(\widehat{\mathbf{R}}_{j,l} \mathbf{W}_l (\widehat{\mathbf{h}}_{k,l}+\widetilde{\mathbf{h}}_{k,l})(\widehat{\mathbf{h}}_{k,l}^{\mathrm{H}}+\widetilde{\mathbf{h}}_{k,l}^{\mathrm{H}}) \mathbf{W}_l)} }{N_l\left (1+e_{j,l} \right )^2} .
    \end{align}
\end{subequations}
We observe the four channel combinations $(\widehat{\mathbf{h}}_{k,l}+\widetilde{\mathbf{h}}_{k,l})(\widehat{\mathbf{h}}_{k,l}^{\mathrm{H}}+\widetilde{\mathbf{h}}_{k,l}^{\mathrm{H}})=\widehat{\mathbf{h}}_{k,l} \widehat{\mathbf{h}}_{k,l}^{\mathrm{H}} + \widehat{\mathbf{h}}_{k,l} \widetilde{\mathbf{h}}_{k,l}^{\mathrm{H}} + \widetilde{\mathbf{h}}_{k,l} \widehat{\mathbf{h}}_{k,l}^{\mathrm{H}} + \widetilde{\mathbf{h}}_{k,l}\widetilde{\mathbf{h}}_{k,l}^{\mathrm{H}}$ separately. 
Firstly, since the error channel $\widetilde{\mathbf{h}}_{k,l}^{\mathrm{H}}$ is uncorrelated with the estimated channel and does not appear in $\mathbf{W}_l$, it is independent of other variables. 
Therefore, with Lemma \ref{xAx} we have
\begin{equation}\label{inf_Rerr}
    \tr{(\widehat{\mathbf{R}}_{j,l} \mathbf{W}_l (\widetilde{\mathbf{h}}_{k,l} \widetilde{\mathbf{h}}_{k,l}^{\mathrm{H}}) \mathbf{W}_l)}
    \stackrel{\mathcal{N} \rightarrow \infty}{\longrightarrow}
    \frac{\tr{(\widehat{\mathbf{R}}_{j,l} \mathbf{W}_l \widetilde{\mathbf{R}}_{k,l}  \mathbf{W}_l)}}{N_l}.
\end{equation}

Similarly, for the case with independent error channels and estimated channels, with Eq. \eqref{QBaIQPsi}, Lemma \ref{xAx} and Lemma \ref{xAy} we have
\begin{subequations}
    \begin{align}
         & \tr{(\widehat{\mathbf{R}}_{j,l} \mathbf{W}_l (\widehat{\mathbf{h}}_{k,l} \widetilde{\mathbf{h}}_{k,l}^{\mathrm{H}}) \mathbf{W}_l)}
        = \tr{( \widetilde{\mathbf{h}}_{k,l}^{\mathrm{H}} \mathbf{W}_l \widehat{\mathbf{R}}_{j,l} \mathbf{W}_l \widehat{\mathbf{h}}_{k,l})}\\
         & = \frac{\tr{( \widetilde{\mathbf{x}}_{k,l}^{\mathrm{H}} \widetilde{\mathbf{R}}_{k,l}^{1/2} \mathbf{W}_l \widehat{\mathbf{R}}_{j,l} \mathbf{A}_{[k],l} \widehat{\mathbf{R}}_{k,l}^{1/2} \widehat{\mathbf{x}}_{k,l})}}{1+e_{k,l}}
        \stackrel{\mathcal{N} \rightarrow \infty}{\longrightarrow} 0.
    \end{align}
\end{subequations}
When we observe $\tr{\left (\widehat{\mathbf{R}}_{k,l} \mathbf{A}_{[k],l}^{-1} \widehat{\mathbf{R}}_{j,l} \mathbf{A}_{[k],l}^{-1} \right )}$, we have
\begin{equation}
    \partial\mathbf{W}_{l,\widehat{\mathbf{R}}_{k,l}}(z)=\mathbf{W}_{l,\widehat{\mathbf{R}}_{k,l}}(z) \widehat{\mathbf{R}}_{k,l} \mathbf{W}_{l,\widehat{\mathbf{R}}_{k,l}}(z)
\end{equation}
Due to the presence of two coherent random matrix $\mathbf{W}_{l}$ in the above expression, further processing is required.
First, let $\mathbf{W}_{l,\widehat{\mathbf{R}}_{k,l}}(z)=(\sum_{k=1}^{K} \mathbf{h}_{k,l}\mathbf{h}_{k,l}^{\mathrm{H}}+\alpha \mathbf{I}_{N_{l}}-z\widehat{\mathbf{R}}_{k,l})^{-1}$, then we have
$\partial\mathbf{W}_{l,\widehat{\mathbf{R}}_{k,l}}(z)=\mathbf{W}_{l,\widehat{\mathbf{R}}_{k,l}}(z) \widehat{\mathbf{R}}_{k,l} \mathbf{W}_{l,\widehat{\mathbf{R}}_{k,l}}(z)$.
Obviously, when $z=0$, we have $\mathbf{W}_{l,\widehat{\mathbf{R}}_{k,l}}(z) \widehat{\mathbf{R}}_{k,l} \mathbf{W}_{l,\widehat{\mathbf{R}}_{k,l}}(z)=\mathbf{W}_{l} \widehat{\mathbf{R}}_{k,l} \mathbf{W}_{l}$.
Therefore, we obtain$\tr{\left (\widehat{\mathbf{R}}_{j,l} \mathbf{W}_{l} \widehat{\mathbf{R}}_{k,l} \mathbf{W}_{l} \right )}=\tr{\left (\widehat{\mathbf{R}}_{j,l} \partial\mathbf{W}_{l,\widehat{\mathbf{R}}_{k,l}}(0) \right )}$.
Similarly, we define
\begin{equation}
    \boldsymbol{\Psi}_{l,\widehat{\mathbf{R}}_{k,l}}(z)=\left ( \frac{1}{N_l}\sum_{i=1}^{K}\frac{s_{i,l}}{1+e_{i,l,\widehat{\mathbf{R}}_{k,l}}}\widehat{\mathbf{R}}_{i,l}+\alpha\mathbf{I}_{N}-z\widehat{\mathbf{R}}_{k,l} \right )^{-1}
\end{equation}
\begin{equation}
    e_{j,l,\widehat{\mathbf{R}}_{k,l}}=\frac{1}{N_l}\tr{(\widehat{\mathbf{R}}_{j,l}\boldsymbol{\Psi}_{l,\widehat{\mathbf{R}}_{k,l}}(z))}
\end{equation}
According to the above definition, we can find its derivative eq. \eqref{def_dPsi_lRkl} at the top of pp. \pageref{def_dPsi_lRkl}.
Furthermore, we can directly derive eq. \eqref{trdef_dPsi_lRkl} at the top of pp. \pageref{trdef_dPsi_lRkl}.
\begin{figure*}
    \begin{equation}\label{def_dPsi_lRkl}
         \partial\boldsymbol{\Psi}_{l,\widehat{\mathbf{R}}_{k,l}}(z)  =
        -\boldsymbol{\Psi}_{l,\widehat{\mathbf{R}}_{k,l}}(z)
        \partial(\boldsymbol{\Psi}_{l,\widehat{\mathbf{R}}_{k,l}}(z))^{-1}
        \boldsymbol{\Psi}_{l,\widehat{\mathbf{R}}_{k,l}}(z)
         =\boldsymbol{\Psi}_{l,\widehat{\mathbf{R}}_{k,l}}(z)
        \left ( \frac{1}{N_l}\sum_{i=1}^{K}\frac{\frac{s_{i,l}}{N_l}\tr{(\widehat{\mathbf{R}}_{i,l}\partial\boldsymbol{\Psi}_{l,\widehat{\mathbf{R}}_{k,l}}(z))}}{(1+e_{i,l,\widehat{\mathbf{R}}_{k,l}})^2}\widehat{\mathbf{R}}_{i,l}+\widehat{\mathbf{R}}_{k,l} \right )
        \boldsymbol{\Psi}_{l,\widehat{\mathbf{R}}_{k,l}}(z)
\end{equation}
\noindent\rule[0.25\baselineskip]{\textwidth}{0.8pt}
\end{figure*}

\begin{figure*}
    \begin{equation}\label{trdef_dPsi_lRkl}
          \frac{1}{N_l}\tr{(\widehat{\mathbf{R}}_{j,l}\partial\boldsymbol{\Psi}_{l,\widehat{\mathbf{R}}_{k,l}}(z))} =\sum_{i=1}^{K}\frac{s_{i,l}\tr{(\widehat{\mathbf{R}}_{i,l}\partial\boldsymbol{\Psi}_{l,\widehat{\mathbf{R}}_{k,l}}(z))}}{N_{l}^3(1+e_{i,l,\widehat{\mathbf{R}}_{k,l}})^2}
          \tr{(\widehat{\mathbf{R}}_{j,l}\boldsymbol{\Psi}_{l,\widehat{\mathbf{R}}_{k,l}}(z) \widehat{\mathbf{R}}_{i,l} \boldsymbol{\Psi}_{l,\widehat{\mathbf{R}}_{k,l}}(z))}
        +\frac{1}{N_l}\tr{(\widehat{\mathbf{R}}_{j,l} \boldsymbol{\Psi}_{l,\widehat{\mathbf{R}}_{k,l}}(z) \widehat{\mathbf{R}}_{k,l} \boldsymbol{\Psi}_{l,\widehat{\mathbf{R}}_{k,l}}(z))}
\end{equation}
\noindent\rule[0.25\baselineskip]{\textwidth}{0.8pt}
\end{figure*}

At this point, if we set the formula as Eq. \eqref{ejvkl}.
Then we can obtain the equation $\mathbf{e}'_{:,l,\widehat{\mathbf{R}}_{k,l}}=\mathbf{J}_{l} \mathbf{e}'_{:,l,\widehat{\mathbf{R}}_{k,l}}+\mathbf{v}_{k,l}$, which can be directly solved by $\mathbf{e}'_{:,l,\widehat{\mathbf{R}}_{k,l}}=(\mathbf{I}_{N_l}-\mathbf{J}_{l})^{-1} \mathbf{v}_{k,l}$. 
Thus, we obtain
\begin{equation}
    \frac{1}{N_l}\tr{\left (\widehat{\mathbf{R}}_{j,l} \mathbf{W}_{l} \widehat{\mathbf{R}}_{k,l} \mathbf{W}_{l} \right )}
    \stackrel{\mathcal{N} \rightarrow \infty}{\longrightarrow}
    \frac{1}{N_l}\tr{(\widehat{\mathbf{R}}_{j,l}\partial\boldsymbol{\Psi}_{l,\widehat{\mathbf{R}}_{k,l}}(0))}.
\end{equation}
Therefore, from \eqref{hwhhwh}, we have the  deterministic equivalent of coherent terms of interference Eq. \eqref{de_co_intf} at the top of pp. \pageref{de_co_intf}.
\begin{figure*}
    \begin{equation}\label{de_co_intf}
         \mathbf{h}_{k,l}^{\mathrm{H}} \mathbf{W}_{l} \widehat{\mathbf{h}}_{j,l}   \widehat{\mathbf{h}}_{j,l}^{\mathrm{H}} \mathbf{W}_{l} \mathbf{h}_{k,l} 
         \stackrel{\mathcal{N} \rightarrow \infty}{\longrightarrow}\frac{1}{N_l(1+e_{j,l})^2} \left ( \frac{e'_{k,l,\widehat{\mathbf{R}}_{j,l}}}{(1+e_{k,l})^2}  + e'_{j,l,\widetilde{\mathbf{R}}_{k,l}}  \right )
    \end{equation}
    \noindent\rule[0.25\baselineskip]{\textwidth}{0.8pt}
\end{figure*}
In the above equation, $e'_{j,l,\widetilde{\mathbf{R}}_{k,l}} = \frac{1}{N_l} \tr{(\widehat{\mathbf{R}}_{j,l} \partial\boldsymbol{\Psi}_{l,\widetilde{\mathbf{R}}_{k,l}})} = \frac{1}{N_l} \tr{(\widetilde{\mathbf{R}}_{k,l} \partial\boldsymbol{\Psi}_{l,\widehat{\mathbf{R}}_{j,l}})}$ can also be calculated according to equation \eqref{def_dPsi_lRkl}.
Due to the utilization of $e_{k,l}$, the above equation is equally applicable to scenarios where $k \notin l$.
\subsection*{Deterministic Equivalent of Normalized Power}
This section solves for the  deterministic equivalent of $\widehat{\mathbf{h}}_{k,l}^{\mathrm{H}} \mathbf{W}_{l}^{\mathrm{H}} \mathbf{W}_{l} \widehat{\mathbf{h}}_{k,l}$. 
First, we have
\begin{subequations}
    \begin{align}
         & \mathbf{h}_{k,l}^{\mathrm{H}} \mathbf{W}_{l} \mathbf{W}_{l} \mathbf{h}_{k,l} \\
         & =\frac{\mathbf{h}_{k,l}^{\mathrm{H}} \mathbf{A}_{[k],l}^{-1}
         \mathbf{A}_{[k],l}^{-1} \mathbf{h}_{k,l}}{\left(1+\mathbf{h}_{k,l}^{\mathrm{H}}\mathbf{A}_{[k],l}^{-1}\mathbf{h}_{k,l}\right)\left( 1+\mathbf{h}_{k,l}\mathbf{A}_{[k],l}^{-1}\mathbf{h}_{k,l}^{\mathrm{H}} \right)} \\
         & \stackrel{\mathcal{N} \rightarrow \infty}{\longrightarrow}
        \frac{1}{\left ( 1+\frac{1}{N_l} \tr{(\widehat{\mathbf{R}}_{k,l}\boldsymbol{\Psi}_l)} \right )^2} \frac{1}{N_l} \tr{\left( \widehat{\mathbf{R}}_{k,l}\mathbf{W}_{l}^2 \right)}
    \end{align}
\end{subequations}
We define $\mathbf{W}_{l}(\alpha)=(\sum_{k}^{K} \mathbf{h}_{k,l}\mathbf{h}_{k,l}^{\mathrm{H}}+\alpha \mathbf{I}_{N_{l}})^{-1}$, and then we have $\partial\mathbf{W}_{l}(\alpha)=\mathbf{W}_{l}(\alpha)^2=\mathbf{W}_{l}^2$.
Referring to the derivation in the previous section, we have
\begin{subequations}
    \begin{align}
        &\partial\boldsymbol{\Psi}_{l}(\alpha)  =
        -\boldsymbol{\Psi}_{l}(\alpha)
        \partial(\boldsymbol{\Psi}_{l}(\alpha))^{-1}
        \boldsymbol{\Psi}_{l}(\alpha)\\
        & =\boldsymbol{\Psi}_{l}(\alpha)
        \left ( \frac{1}{N_l}\sum_{i=1}^{K}\frac{\frac{s_{i,l}}{N_l}\tr{(\widehat{\mathbf{R}}_{i,l}\partial\boldsymbol{\Psi}_{l}(\alpha))}}{(1+e_{i,l})^2}\widehat{\mathbf{R}}_{i,l}+\mathbf{I}_{N_{l}} \right )
        \boldsymbol{\Psi}_{l}(\alpha)
    \end{align}
\end{subequations}
It is not difficult for us to set the formula as Eq. \eqref{ejvl}, then we obtain $\mathbf{e}'_{:,l}=\mathbf{J}_{l} \mathbf{e}'_{:,l}+\mathbf{v}_{l}$. 
Thus, we have the  deterministic equivalent of normalized power eq. \eqref{de_nor_pow} at the top of pp. \pageref{de_nor_pow}.
\begin{figure*}
\begin{equation}\label{de_nor_pow}
    1/\rho_{k,l}=\mathbf{h}_{k,l}^{\mathrm{H}} \mathbf{W}_{l} \mathbf{W}_{l} \mathbf{h}_{k,l}
    \stackrel{\mathcal{N} \rightarrow \infty}{\longrightarrow}
    \frac{1}{\left ( 1+\frac{s_{i,l}}{N_l} \tr{(\widehat{\mathbf{R}}_{k,l}\boldsymbol{\Psi}_l)} \right )^2} \frac{1}{N_l}\tr{(\widehat{\mathbf{R}}_{k,l}\partial\boldsymbol{\Psi}_{l})}
    =\frac{e'_{k,l}}{(1+e_{k,l})^2}.
\end{equation}
\noindent\rule[0.25\baselineskip]{\textwidth}{0.8pt}
\end{figure*}

\subsection{Important Lemmas}
From \cite{Wagner2012TIT}, we obtain the following theorem \ref{Psi_e_k} and lemma \ref{xAx}, \ref{xAy} and \ref{B-Bvv}:
\begin{theorem}\label{Psi_e_k}
    For an $N \times N$ matrix of the form
    \begin{equation}
    \mathbf{B}=\sum_{k=1}^{K}\mathbf{R}_{k}^{\frac{1}{2}}\mathbf{x}_{k}\mathbf{x}_{k}^{\mathrm{H}}\mathbf{R}_{k}^{\frac{1}{2}} + \mathbf{S} + \alpha\mathbf{I},
    \end{equation}
    where $\mathbf{R}_{k}$ are symmetric positive definite matrices, $\mathbf{R}_{k}^{\frac{1}{2}}$ is a decomposition satisfying $\mathbf{R}_{k}=\mathbf{R}_{k}^{\frac{1}{2}}(\mathbf{R}_{k}^{\frac{1}{2}})^{\mathrm{H}}$ (e.g., Cholesky decomposition), $\mathbf{S}$ is a Hermitian non-negative definite matrix, $\mathbf{x}_{k} \sim \mathcal{CN}(0,\frac{1}{N}\mathbf{I}_{N})$, and $\mathbf{I}_{N}$ is an $N \times N$ identity matrix.
    
    As $\mathcal{N} \rightarrow \infty$, we have
    \begin{equation}
        \frac{1}{N}\tr{(\mathbf{Q}(\mathbf{B}+\alpha\mathbf{I}_{N})^{-1})}
        - \frac{1}{N}\tr{(\mathbf{Q}\boldsymbol{\Psi})}
        \stackrel{\mathcal{N} \rightarrow \infty}{\longrightarrow} 0,
    \end{equation}
    where the spectral norm of $\mathbf{Q}$ is uniformly bounded, and
    \begin{equation}\label{Psi_inv}
        \boldsymbol{\Psi}=\left ( \frac{1}{N}\sum_{i=1}^{K}\frac{1}{1+e_i}\mathbf{R}_{i}+\mathbf{S}+\alpha\mathbf{I}_{N}\right )^{-1}
    \end{equation}
    \begin{equation}\label{e_tr}
        e_{k}=\frac{1}{N}\tr(\mathbf{R}_k\boldsymbol{\Psi})
    \end{equation}
    
    Here, $\alpha \in \mathbb{R}^{+}$. Equations \eqref{Psi_inv} and \eqref{e_tr} form a fixed-point equation that can be solved iteratively.
\end{theorem}

\begin{lemma}\label{xAx}
    For $\mathbf{x}_{k} \sim \mathcal{CN}(0,\frac{1}{N}\mathbf{I}_{N})$, $\mathbf{I}_{N}$, and another independent $N \times N$ random matrix $\mathbf{A}$, we have
    \begin{equation}
        \mathbf{x}_{k}^{\mathrm{H}}\mathbf{A}\mathbf{x}_{k}-\frac{1}{N}\tr(\mathbf{A})
        \stackrel{\mathcal{N} \rightarrow \infty}{\longrightarrow} 0
    \end{equation}
\end{lemma}

\begin{lemma}\label{xAy}
    For $\mathbf{x}_{k} \sim \mathcal{CN}(0,\frac{1}{N}\mathbf{I}_{N})$, $\mathbf{I}_{N}$, $\mathbf{y}_{k} \sim \mathcal{CN}(0,\frac{1}{N}\mathbf{I}_{N})$, $\mathbf{I}_{N}$, and an $N \times N$ finite random matrix $\mathbf{A}$, where $\mathbf{x}_{k}$, $\mathbf{y}_{k}$, and $\mathbf{A}$ are mutually independent, we have
    \begin{equation}
        \mathbf{x}_{k}^{\mathrm{H}}\mathbf{A}\mathbf{y}_{k}\stackrel{\mathcal{N} \rightarrow \infty}{\longrightarrow} 0
    \end{equation}
\end{lemma}

\begin{lemma}\label{B-Bvv}
    For any $\mathbf{A} \in \mathbb{C}^{N \times N}$ be deterministic with uniformly bounded spectral norm and $\mathbf{B}$ be random Hermitian, with eigenvalues $\lambda_1 \leq \lambda_2 \leq \cdots \leq \lambda_N$ such that, with probability $1$, there exist $\epsilon \geq 0$ for which $\lambda_1 \geq \epsilon$ for any large $N$.
    Then for $\mathbf{v} \in \mathbb{C}^{N \times 1}$, we have
    \begin{equation}
        \frac{1}{N}\tr{(\mathbf{A}\mathbf{B}^{-1})}-\frac{1}{N}\tr{(\mathbf{A}(\mathbf{B}+\mathbf{v}\mathbf{v}^{\mathrm{H}})^{-1})}
        \stackrel{\mathcal{N} \rightarrow \infty}{\longrightarrow} 0
    \end{equation}
\end{lemma}

From \cite{Bai2009WS}, we obtain
\begin{lemma}\label{qAq}
    For any invertible matrix $\mathbf{A} \in \mathbb{C}^{n \times n}$ and $\mathbf{q} \in \mathbb{C}^{n \times 1}$, we have
    \begin{equation}
        \mathbf{q}^{\mathrm{H}}\left ( \mathbf{A} + \mathbf{q}\mathbf{q}^{\mathrm{H}} \right )^{-1}=\frac{1}{1+\mathbf{q}^{\mathrm{H}}\mathbf{A}^{-1}\mathbf{q}}\mathbf{q}^{\mathrm{H}}\mathbf{A}^{-1}
    \end{equation}
\end{lemma}

\bibliographystyle{IEEEtran}

\bibliography{ref}

\end{document}